\def\insfig#1{#1}
\def\endinsfig{\end{document}}

\def\manuscript{
	\documentstyle[12pt,aasms4]{article}
	\def\insfig##1{}
	\def\endinsfig{}
	\def\manufig##1{}   
	\def\capfig##1{##1}  
	}

\def\typeset{
	\documentstyle[12pt,aasms4]{article}
	\def\insfig##1{}
	\def\endinsfig{}
	\def\manufig##1{##1} 
	\def\capfig##1{}    
	}

  \documentstyle[11pt,aas2pp4]{article}  

\font\smallrm=cmr8

\def\PD{\hbox{\smallrm PD}}
\def\RMS{{\smallrm RMS}}

\def\kms{\hbox{$\,$km$\,$s$^{-1}$}}

\def\etal{{et~al.}}

\def\mi{\ifmmode\overline{m}_I\else$\overline{m}_I$\fi}
\def\mv{\ifmmode\overline{m}_V\else$\overline{m}_V$\fi}
\def\Mi{\ifmmode\overline{M}_I\else$\overline{M}_I$\fi}
\def\Miz{\ifmmode\overline{M}_I^0\else$\overline{M}_I^0$\fi}
\def\mim{\ifmmode\overline{m}_I^0\else$\overline{m}_I^0$\fi}
\def\dmod{\ifmmode(m{-}M)\else$(m{-}M)$\fi}
\def\vi{\ifmmode(V{-}I)\else$(V{-}I)$\fi}
\def\viz{\ifmmode(V{-}I)_0\else$(V{-}I)_0$\fi}
\def\dn{\ifmmode D_n{-}\sigma\else$ D_n{-}\sigma$\fi}
\def\avemi{\ifmmode\langle\overline{m}_I^0\rangle\else$\langle\overline{m}_I^0\rangle$\fi}
\def\Nbar{\ifmmode\overline{N}\else$\overline{N}$\fi}
\def \gap
 {\mathrel{\hbox{\raise0.3ex\hbox{$>$}\kern-0.8em\lower0.8ex\hbox{$\sim$}}}}
\def \lap
 {\mathrel{\hbox{\raise0.3ex\hbox{$<$}\kern-0.75em\lower0.8ex\hbox{$\sim$}}}}

\lefthead{Tonry, Dressler, Blakeslee, and Ajhar}
\righthead{SBF Distances}

\begin{document}

\title{The SBF Survey of Galaxy Distances. IV. \\
     SBF Magnitudes, Colors, and Distances\altaffilmark{1}}

\author{John L. Tonry\altaffilmark{2}}
\affil{Institute for Astronomy, University of Hawaii, Honolulu, HI 96822}
\affil{Electronic mail: jt@ifa.hawaii.edu}
\authoremail{jt@ifa.hawaii.edu}

\author{Alan Dressler}
\affil{Carnegie Observatories, 813 Santa Barbara St., Pasadena, CA 91101}
\authoremail{dressler@omega.ociw.edu}
 
\author{John P. Blakeslee\altaffilmark{3}}
\affil{Dept.\ of Physics, University of Durham, South Road, Durham, DH1 3LE,
United Kingdom}
\authoremail{j.p.blakeslee@durham.ac.uk}
 
\author{Edward A. Ajhar\altaffilmark{2}}
\affil{Kitt Peak National Observatory, National Optical Astronomy
Observatories, P. O. Box 26732,}
\affil{Tucson, AZ 85726}
\authoremail{ajhar@noao.edu}
 
\author{Andr\'{e} B. Fletcher}
\affil{MIT Haystack Observatory, Off Route 40, Westford, MA 01886}

\author{Gerard A. Luppino}
\affil{Institute for Astronomy, University of Hawaii, Honolulu, HI 96822}

\author{Mark R. Metzger}
\affil{Caltech Astronomy Dept, MS\,105-24, Pasadena, CA 91125}

\author{Christopher B. Moore}
\affil{Harvard-Smithsonian Center for Astrophysics,
60 Garden St., Cambridge, MA 02140}


\altaffiltext{1}{Observations in part from the Michigan-Dartmouth-MIT 
(MDM) Observatory.}
\altaffiltext{2}{Guest observers at the Cerro Tololo Inter-American
Observatory and the Kitt Peak National Observatory, National Optical
Astronomy Observatories, which are operated by AURA, Inc., under
cooperative agreement with the National Science Foundation.}
\altaffiltext{3}{Current address: Department of Physics and Astronomy,
Johns Hopkins University, Baltimore, MD 21218}

\begin{abstract}

We report data for $I$ band Surface Brightness Fluctuation (SBF)
magnitudes, \vi\ colors, and distance moduli for 300 galaxies.  The
Survey contains E, S0 and early-type spiral galaxies in the proportions
of 49:42:9, and is essentially complete for E galaxies to Hubble
velocities of 2000\kms, with a substantial sampling of E galaxies out
to 4000\kms.  The median error in distance modulus is 0.22 mag.

We also present two new results from the Survey. (1) We compare the
mean peculiar flow velocity (bulk flow) implied by our distances with
predictions of typical cold dark matter transfer functions as a
function of scale, and find very good agreement with cold, dark matter
cosmologies if the transfer function scale parameter $\Gamma$, and the
power spectrum normalization $\sigma_8$ are related by $\sigma_8
\Gamma^{-0.5} \approx 2\pm0.5$.  Derived directly from velocities, this
result is independent of the distribution of galaxies or models for biasing.
The modest bulk flow contradicts reports of large-scale, large-amplitude
flows in the $\sim200$ Mpc diameter volume surrounding our Survey
volume.  (2) We present a distance-independent measure of absolute
galaxy luminosity, \Nbar, and show how it correlates with galaxy
properties such as color and velocity dispersion, demonstrating its
utility for measuring galaxy distances through large and unknown
extinction.

\end{abstract}

\keywords{galaxies: distances and redshifts ---
galaxies: clusters: individual (Virgo, Centaurus) -- 
cosmology: distance scale --
cosmology: large-scale structure of universe}

\section{Introduction: A Brief History of SBF}

The $I$-band Surface Brightness Fluctuation (SBF) Survey was undertaken
to measure accurate distances to nearby galaxies in the expectation of
tying Cepheid distances to the far-field Hubble flow and substantially
improving our knowledge of the local velocity field.  The Survey was
inaugurated with the work of Tonry, Ajhar, \& Luppino (1990), who
measured $VRI$ SBF magnitudes with the Kitt Peak 4\,m telescope for 14
early-type galaxies, mostly members of the Virgo cluster.  Tonry (1991)
followed with the first fully empirical, though provisional,
calibration of the method, based on the Cepheid distance to M31 and the
color dependence of the $I$-band SBF magnitude \mi\ for Fornax cluster
galaxies observed with the 4\,m telescope at Cerro Tololo.  The data
collection has spanned a decade now, with most of the recent northern
data coming from the MDM 2.4\,m on Kitt Peak and the southern data
coming from the Las Campanas 2.5\,m.  A recent review of SBF can be
found in Blakeslee, Ajhar, \& Tonry (1999).

The present series of papers began with Tonry \etal\ (1997, hereafter
SBF-I), which detailed how the SBF survey data from different
telescopes and observing runs were intercompared and brought into a
homogeneous system, and how error estimates were derived.  In SBF-I, a
new $I$-band calibration with a significantly different zero point was
presented, and various methods were used to tie to the far-field Hubble
flow in order to derive the Hubble constant. Tonry \etal\ (2000,
SBF-II) used the SBF survey data to construct a parametric flow model
which included infall into the Virgo and the Great Attractors, a
residual quadrupole, an overall dipole (bulk flow of the sample), the
cosmic thermal velocity dispersion, and the Hubble constant, all as
free parameters.  SBF-II also revised the zero point by using the
latest Cepheid distances tabulated by Ferrarese \etal\ (2000) and the
new DIRBE/IRAS Galactic extinction estimates from Schlegel, Finkbeiner,
\& Davis (1998).  Blakeslee \etal\ (1999, SBF-III) used the same data
set to compare the measured SBF peculiar velocities to predictions that
derived from the galaxy density field measured by flux-limited redshift
surveys, under the assumption that light traces mass with a linear
biasing prescription.  This comparison yielded values for both the
Hubble constant and $\beta\equiv\Omega^{0.6}/b$, where $\Omega$ is the
matter density and $b$ is the linear bias of the observed galaxies.

In addition to the $I$-band ground-based survey, there have been a
number of recent SBF efforts in the F814W filter of WFPC2 on the Hubble
Space Telescope (HST).  Ajhar \etal\ (1997) gave the initial calibration
of the method for this filter.  Lauer \etal\ (1998) measured SBF
distances to four brightest cluster ellipticals at distances of
$\sim\,$5000\kms, and Pahre \etal\ (1999) measured the distance to
NGC\,4373 in the Hydra-Centaurus supercluster.

Recently, SBF measurements in near-infrared bandpasses have been
added.  Jensen, Luppino, \& Tonry (1998a, 1998b) measured $K$-band SBF
distances to galaxies in the Virgo, Fornax, Eridanus, Centaurus, and
Coma clusters.  Another HST study (Jensen \etal\ 2000, in preparation)
calibrates the F160W near-infrared NICMOS filter for use with SBF and
samples the Hubble flow out to 10,000\kms.  The use of SBF in stellar
population synthesis has received renewed interest of late (e.g. Liu,
Charlot, \& Graham 2000; Blakeslee, Vazdekis, \& Ajhar 2000), both as a
means for probing stellar populations and as an independent check of
the Cepheid distance scale.  These efforts underscore the importance of
a solid empirical calibration of the method.

In this paper, we present and discuss the data from our $I$-band SBF 
distance survey and present two new results based on these
data.

\section{The Galaxy Sample}

Because it was our intent to make this survey as homogeneous as
possible our observations span the entire sky, except for incomplete
sampling of the zone of avoidance.  It is difficult to make a really
quantitative statement of our selection function, because SBF
observability depends strongly on the atmospheric seeing, and we were
quite limited by the availability of telescope time at the many
facilities that were used to conduct this study.  However, as a guide,
we can compare with the RC3: Figure 1 shows how the cumulative counts
rise with redshift in the SBF sample and the RC3.  These are counts of
galaxies with $B_T \le 12.5$, which includes about 63\% of the galaxies
in the SBF survey.  We are nearly complete in E galaxies to a Hubble
velocity of 2000\kms\ and have sampled a very substantial fraction even
to 4000\kms.  The higher incompleteness for S0's compared to E's
reflects both the added difficulty of adequate modeling of disk + bulge
for some S0 galaxies and the fact that we chose not to observe many of
the S0's in groups where we already had distances to elliptical
galaxies.  We have succeeded in measuring SBF magnitudes in only a few
spiral galaxies, mostly ones with large, smooth bulges, and they are
mostly at nearby distances where Cepheid observations are possible.
\insfig{
\begin{figure}[t]
\epsscale{1.0}
\plotone{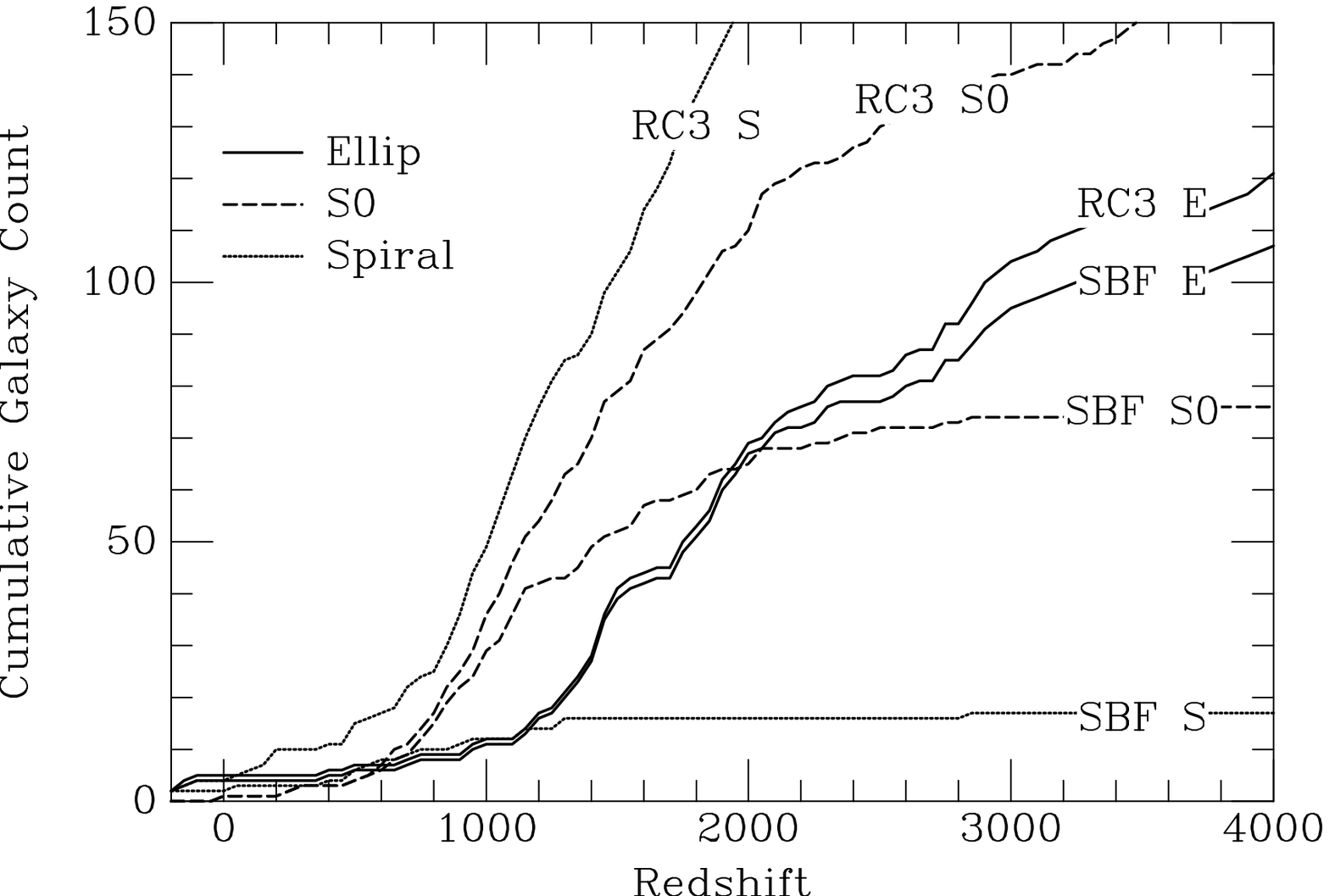}
\caption[rc3lookup.eps]{
The cumulative counts of galaxies with $B_T \le 12.5$ are shown for
the RC3 (upper curves) and the SBF survey (lower curves).  The three
sets of curves show counts of E, S0, and Sa and Sb spiral galaxies
with $T\le3$.
\label{fig:complete}}
\end{figure}
}

\insfig{
\begin{figure}[t]
\epsscale{1.0}
\plotone{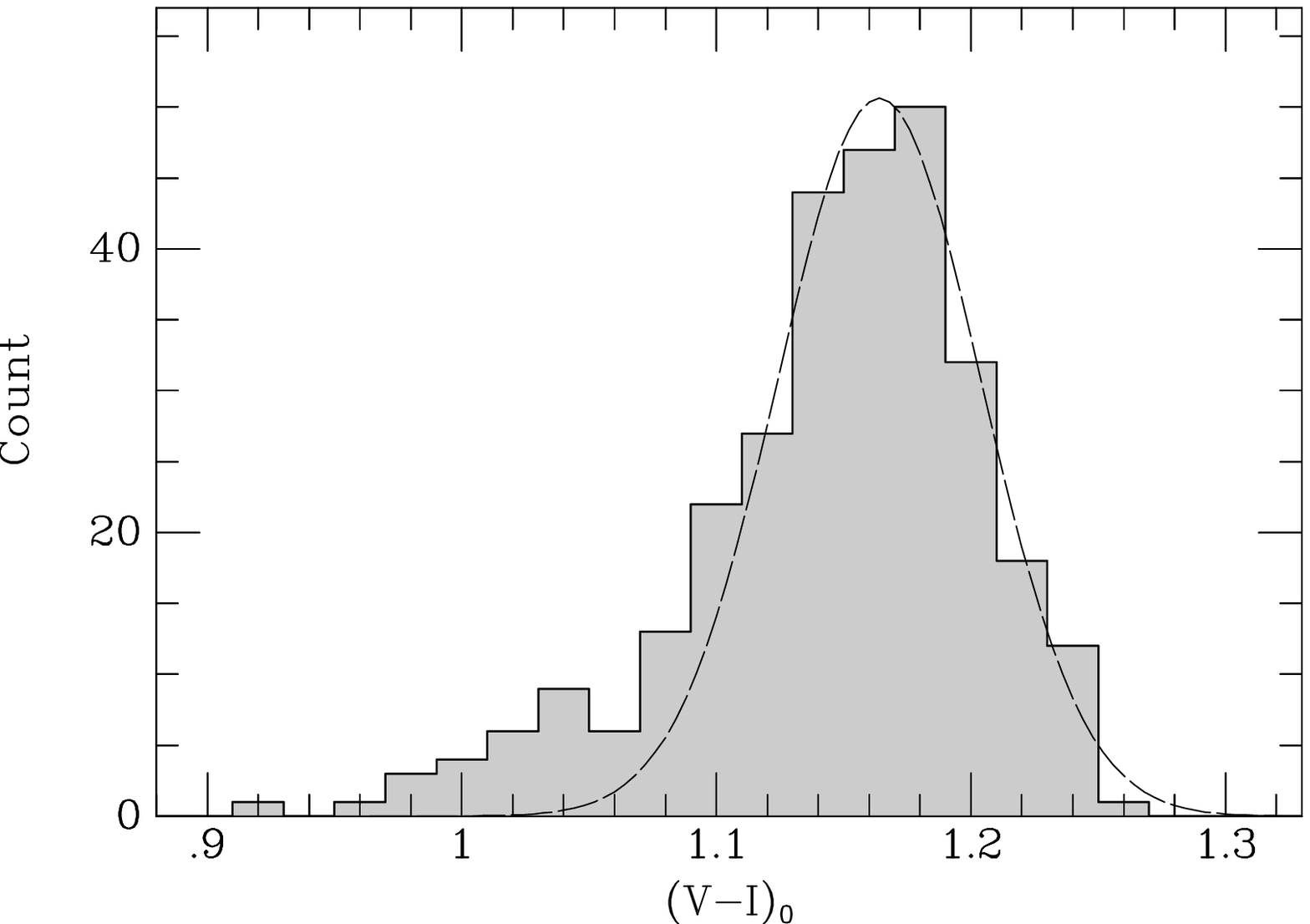}
\caption[vihisto.eps]{
The \vi\ distribution of SBF survey galaxies, represented by the
shaded histogram.  The dashed curve shows for comparison a Gaussian
of mean $\vi=1.16$ and dispersion 0.04\,mag.
\label{fig:vihisto}}
\end{figure}
}

Figure~\ref{fig:vihisto} displays the distribution of galaxies in
reddening-corrected \vi\ color.  The colors used here (those in
Table~\ref{tab:1}) are derived from the regions of the galaxies in
which the SBF analyses were performed.  Overall, the mean color is
$\langle{V{-}I}\rangle_0=1.145$ and the rms dispersion is 0.06\,mag.
The ellipticals follow a more or less Gaussian color distribution
centered at $\viz=1.165$ with a dispersion of only 0.04\,mag.  The blue
tail extending to $\vi<1.0$ is comprised mainly of S0 galaxies.

\section{Calibration Issues}

We described in SBF-I how we brought all of the photometry onto a
common system and how we validated our error estimates for \vi\ and \mi\ 
through multiple observations.   Groups of galaxies were used
to establish the slope of the \mi--\vi\ relationship, and we
compared the relative group distances with those from other distance
estimators to demonstrate the apparent universality of the \mi--\vi\ relationship.  By comparison with the extant Cepheid distances we also
chose a zero point for the \Mi--\vi\ relation and reported that the 
relations for ellipticals, S0 galaxies and spiral bulges are 
indistinguishable with the present data.

In SBF-II, we described our switch from the older Burstein \& Heiles
(1984) H{$\,$\sc i}-derived extinction estimates to the new Schlegel,
Finkbeiner, \& Davis (1998, hereafter SFD) extinctions based on the
DIRBE/IRAS maps.  The SFD values are preferable because of their
greater homogeneity over the sky and greater angular resolution, and
they are probably more accurate (see SFD).  The change affects our
reductions of both \mi\ and \vi\ to extinction-free values.  Despite an
overall shift in $E(B{-}V)$ of nearly 0.02 mag, there was little change
to the distance estimates for most galaxies because our zero point also
comes from galaxies with revised extinction estimates.  Nevertheless,
these values for \mi\ and \vi\ should {\it not} be mixed with
previously published data without due attention to the differing
assumptions about extinction.

The zero point for the \Mi--\vi\ relation continues to be a work in
progress, unfortunately, as we discuss in detail in SBF-II.  While
there is concurrence that the slope of the \Mi--\vi\ relation derived
in SBF-I is accurate, uncertainty in the zero point remains at the 0.1
mag level.  Essentially, the SBF zero point can be calibrated using
Cepheid distances or theory.  The Cepheid distances can be applied to
SBF measurements either galaxy by galaxy or through group association
of galaxies with Cepheid distances to other galaxies with SBF
distances.  SBF-II adopted the former approach and derived
\begin{equation}
  \Mi = -1.74\pm0.08 + (4.5\pm0.25)[\viz - 1.15]
  \label{eq:zeropt}
\end{equation}
using the HST Key Project Cepheid distances (Ferrarese et al. 2000). The
validity of this equation has been tested for colors in the range of
$0.95 < \viz < 1.30$.  Our zero point is based on the median value for
six spirals with both Cepheid and SBF measurements; using a similar
approach and a weighted average, Ferrarese et al. (2000) derived a zero
point of $-1.79\pm0.09$.

The uncertainty in the SBF zero-point is illustrated by comparing these
values to the result of the group-association method. As we discuss in
SBF-II, using a group association between Cepheid-bearing spirals and
early-type galaxies with SBF measurements results in a significant
change in the zero point to $-1.61\pm0.03$.  The fainter group-based
calibration points to either systematically brighter SBF magnitudes for
spiral bulges or, as found by Kelson \etal\ (2000) in the context of
the fundamental plane calibration, a systematic offset for the Key
Project spirals to lie in front of the ellipticals in the same groups.
Additional data will be needed to resolve the cause of this discrepancy
and thereby firm up the SBF calibration.

We emphasize that the error bars given here do {\it not} include the
systematic uncertainty in the Cepheid zero point, estimated by Mould
\etal\ (2000) to be $\pm0.16$ mag, which allows for $\pm0.13$ mag of
uncertainty in the adopted LMC distance modulus of 18.50\,mag.  This
uncertainty should be taken seriously, since the distance to the LMC
remains controversial.  At the time of this writing, Freedman et al. 
(2000) are preparing a major revision in the Cepheid distance scale
based on a reassessment of the LMC distance from seven independent
means, an improved PL relation based on the new, large sample of LMC
Cepheids from the OGLE Project, and a small metallicity dependence
for the PL relation.  Application of these new Cepheid
distances to our six calibrating galaxies will make our SBF calibration
fainter by $\sim0.1$ mag with a corresponding rise in the Hubble
constant of about 5\%.

A final example of the continuing uncertainty in the SBF calibration
comes from its derivation purely from stellar evolution theory and
population synthesis.  Worthey (1993a, 1993b, 1994) was the first to
accurately reproduce the empirical slope of the \Mi--\vi\ relation;
these models gave a theoretical zero point of $-1.81$ (see SBF-I),
consistent with the value in eq.\,(\ref{eq:zeropt}) from the Cepheid
calibration.  Liu \etal\ (2000) use an updated version of the Bruzual
\& Charlot (1993) models with the isochrones of Bertelli \etal\ (1994)
and find a similar $I$-band zero point value of $-1.79$.  
These authors revised the AGB evolution in their models because the
original Bruzual \& Charlot AGB prescription gave a \Mi\ zero point
significantly fainter than the empirical result.
The uncertainty in modeling the AGB is
a lingering problem for the theoretical derivation of the $I$-band
zero point and will be an even bigger problem for
the $K$ band.  Indeed, a fully independent re-derivation
by Blakeslee \etal\ (2000) using new models based on the latest
isochrones from the Padua group (Girardi \etal\ 2000) and empirical
color-temperature transformations has yielded a value that is
significantly different. They reproduce the observed fluctuation colors
well and also match the empirical $I$-band slope,
but find a zero point of $-1.47$, i.e., 0.27\,mag brighter than in
eq.\,(\ref{eq:zeropt}). The discrepancy drops to 0.14\,mag for the
SBF-II group calibration, and then would essentially disappear for
the revised Key Project Cepheid scale referred to above.
As Blakeslee \etal\ discuss, the theoretical uncertainty
also reflects a moderate sensitivity to the uncertain details
of stellar evolution, for example, the distribution and lifetimes 
of stars on the red giant branch.  

Although we look forward to advances in stellar evolution theory that
will allow a robust theoretical SBF zero point, this is not likely to
happen soon.  For now, the empirical zero point appears more
reliable.


\section{The Data}

Table 1 presents colors, fluctuation magnitudes, and distance moduli
of the galaxies in the SBF sample.  The columns are 
(1) galaxy name;
(2) right ascension (J2000) from the RC3; 
(3) declination (J2000); 
(4) redshift (km$\,s^{-1}$, CMB reference frame); 
(5) morphological $T$ type; 
(6) group number as defined by the Faber et al. (1989 7S);
(7) $B$ band extinction adopted from SFD;
(8) $(V{-}I)$ color measured at radii indicated in column 11, error,
and number of contributing measurements; 
(9) $I$ band fluctuation magnitude, error, and number of
contributing measurements; 
(10) distance modulus using eq.\,(\ref{eq:zeropt}) for $\overline M_I$
as a function of $(V{-}I)$, including all sources of error except
systematic error in the zero point;
(11) mean annular radius (arcsec) contributing to the SBF measurement,
 ratio of the innermost contributing radius to the mean, and
 ratio of the outermost contributing radius to the mean; 
(12) observation quality, as defined in SBF-II: $Q =
\log_2[N_e(\overline m)/\PD^2]$, where $N_e(\overline m)$ is the
number of electrons that would be detected  
in the image from an object of magnitude $\overline m$, 
and \PD\ is the product of the full-width at half-maximum of the 
point spread function (PSF) in arcseconds and
the CMB frame velocity in units of 1000 \kms\
(so $\PD^2$ is proportional to the metric area within a resolution element);
(13) ``\PD'' value defined above; and 
(14) $\overline N_I$, discussed below.  
Note that the seeing (in arcseconds) of the 
observation can be derived as \PD\ divided by the CMB velocity in units
of 1000\kms, except that galaxies within 3$^\circ$ of NGC\,4486 and with
heliocentric velocity less than 2400\kms\ were assigned a \PD\ velocity
of 1350\kms, galaxies within 2$^\circ$ of NGC\,4709 and with
heliocentric velocity between 3800 and 4800\kms\ were assigned a 
\PD\ velocity of 3100\kms, and other galaxies with a negative redshift 
were assigned a velocity of 10\kms.

Table 2 adds a few galaxies whose distances are potentially biased
either because the seeing was poor for the distance ($\PD > 2.7$) or
the quality was poor ($Q < 0$).  These observations were deemed
unacceptable for use in SBF-II, but we include them here despite the
potential for bias they do provide at least crude distances and because
the photometry is good.  (The data of Table 2 might also be useful to
further study of the degree of bias present in this data set.)  We also
include NGC\,3413 in this table because it is a very dusty galaxy and
its color is significantly bluer than our empirical calibration allows.
NGC\,4627 has likewise been relegated to this table for being too
blue.  We still lack adequate photometry to provide colors for
NGC~1331, NGC~3522, and IC~5269.  The columns in this table are the
same as with Table 1.  We stress that these data should be used with
caution and should not be mixed with the observations from Table 1.

Tables 1 and 2 are available in digital form from
http://www.ifa.hawaii.edu/$\sim$jt/SBF in table.good and table.poor.

Table 3 gives values for $\overline m$ in different bandpasses.  The
columns are mostly self explanatory, but the second column lists the
mean radius (in arcsec) where the fluctuation magnitudes and colors
were measured in NGC\,205.  These data come from a single observing run
(M0893; see SBF-I for details), and although they are subjected to the
same calibration as the other data, some of the $\overline m_I$ will
not be identical to Table 1.  We hope, however, that the internal
consistency of these data will be better than by using the average
$\overline m_I$ from Table 1 and therefore the fluctuation colors will
be more accurate.  Note that these data make it obvious that NGC\,404
is not a member of the Local Group, but lies at a distance of about
3\,Mpc.

Table 4 gives ``group average'' distances for the galaxy defined in
SBF-I.  In particular, the membership criteria are based on position on
the sky and redshift, not necessarily the group assignment made by the
7S, although we give the group number defined by the 7S that most
closely corresponds to the SBF groups.  The distance errors do not take
any account of the extent of the group, hence are probably
unrealistically small.  It would now probably be appropriate to select
the groups with some knowledge of the SBF distances (this would have
been circular in SBF-I);  what we present here is only meant as a rough 
guide to group properties.  More precise group distances can be derived 
from better group definitions and from Table 1.

\section{The Local Velocity Field}

Using the data just presented, we made a first effort to extract the
peculiar velocities in SBF-II (where we fit a parametrized model) and
SBF-III (where we match the flow field inferred from IRAS galaxy
counts).  The important results from those papers are that we prefer
the value for $H_0$ of $74\pm4$ from SBF-III (but dependent on the
Cepheid zero point as described above).  We find in SBF-III that
$\beta_{I} = \Omega^{0.6}/b_{I} = 0.42$ for the IRAS 1.2 Jy survey
(Fisher et al. 1995), and $\beta_{O} = 0.26$ for the ``Optical Redshift
Survey'' (Santiago et al. 1995), which is consistent with $b_{IRAS} = 1$,
$b_{opt} = 1.6$, and $\Omega_M = 0.25\pm0.05$.

Because of the interest over the last five years in even larger flows
on very large scales (e.g., Lauer \& Postman 1994; Hudson \etal\ 1999;
Willick 1999), we wish to expand here on the result in SBF-II which
contradicts these claims.  Basically, the evidence is that, according to
our SBF model, the major part of the CMB dipole is generated within a
volume $V \lap 5000 \kms$.  This means that to accept the existence of
these larger bulk flows in a volume that is much larger than this (but
includes it) is to find that our subvolume is basically at rest with
respect to the CMB while material around that volume is, on average
moving with a high velocity ($\gap600 \kms$).  This situation 
implies an improbable, if not impossible, spectrum of initial 
fluctuations, very different from the scale-free power spectrum that
underlies cold-dark-matter (CDM) models.

In contrast, our measure of the bulk flow as a function of scale 
within our SBF Survey volume is in good agreement with theoretical 
predictions of typical cold-dark-matter (CDM) models based on a
scale-free power spectrum, as we now show. The ``SBF-II model'' gives 
us a smooth measure of mean velocity and velocity dispersion throughout 
the survey volume, which is valid within $|SGX|,|SGY| < 50$~Mpc, and 
$|SGZ| < 25$~Mpc. Figure \ref{fig:bulky} illustrates how well a typical 
CDM power spectrum matches the mean bulk flows of these data. 
\insfig{
\begin{figure}[t]
\epsscale{1.0}
\plotone{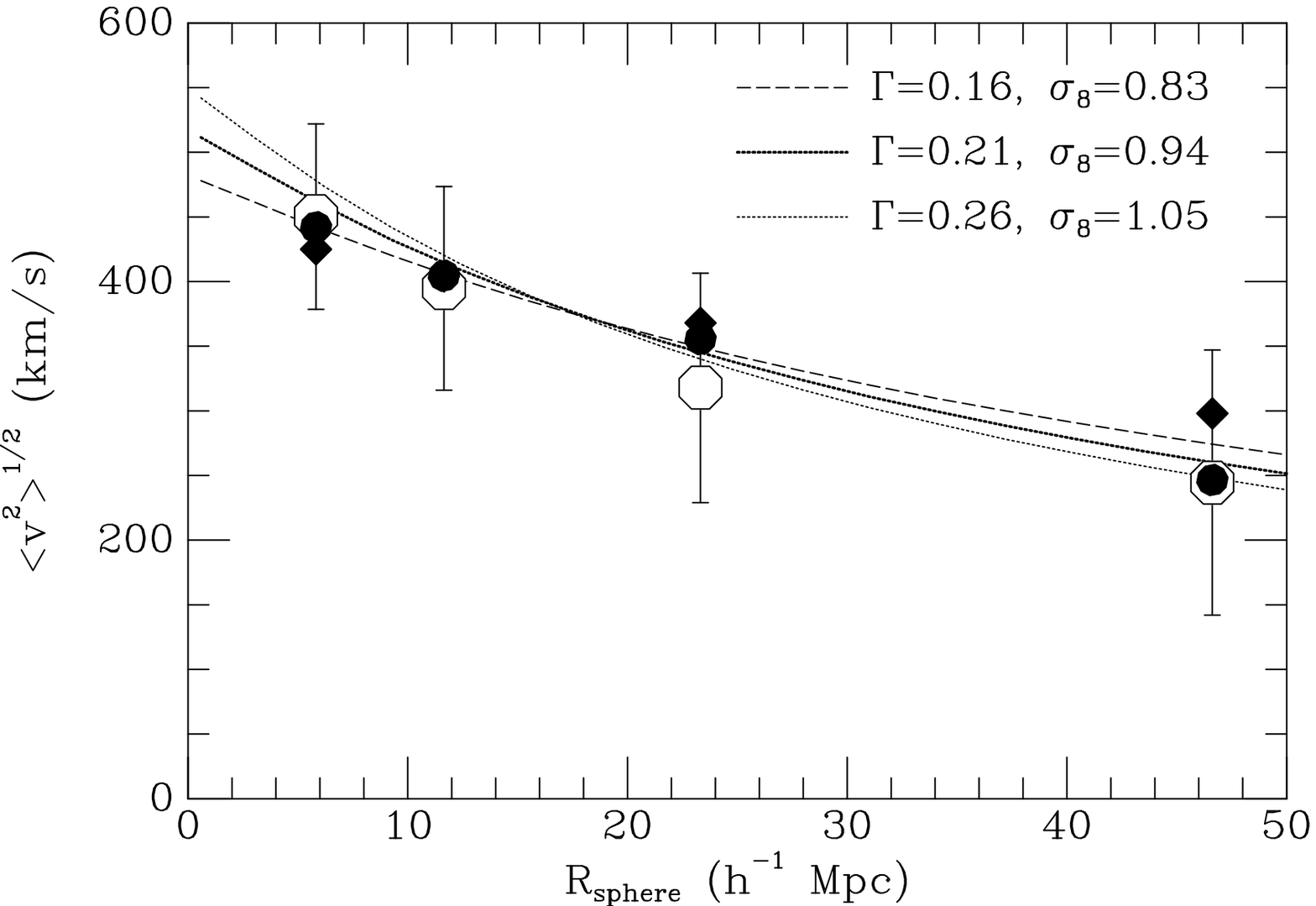}
\caption[bulkflow.eps]{
The bulk flow observed and predicted in top-hat spheres of radius $R$ is
shown for typical CDM models of the power spectrum (curves), a
simulated universe with a $\Gamma=0.21, \sigma_8=0.94$ CDM power
spectrum (large open points), and for the SBF-II observations (circles 
for SBF-II model, diamonds for Willick and Batra variant).  The simulated 
universe encompasses many volumes equal to that of the SBF-II survey, 
and the rms variation in bulk flow seen from volume to volume is shown 
as error bars on the open points.  Note that the radius is expressed 
in terms of $h^{-1}$~Mpc for consistency with what is normally found 
in the literature.  The rollover in the data points at $R<10h^{-1}$~Mpc 
is caused by the limited spatial resolution of the SBF-II model.
\label{fig:bulky}}
\end{figure}
}

The points are generated by computing the SBF-II model velocity with
respect to the CMB, averaged over cubes of varying size.  The RMS of
all the cubes within the survey volume gives us a bulk flow on that
scale.  We assign an effective spherical radius to the cube size by
matching the volume of the sphere to that of the cube.  The rolloff in 
the bulk flow data points at $R<10h^{-1}$~Mpc is inherent in the 
parametrized model.

Willick and Batra (2000) have recently reexamined the SBF survey data
and compared the peculiar velocities with those expected from the IRAS
density distribution.  In addition, they redetermined parameters for
the SBF-II model, and the bulk flows from the Willick and Batra model
are also shown in Figure \ref{fig:bulky} to give a sense of the
uncertainty inherent in the interpolation of the SBF peculiar
velocities by this parametric model.  

For comparison we also plot curves showing the bulk flow expected from a 
typical CDM power spectrum, using the parametrization of Bardeen et al. 
(1986) of the CDM transfer function and a flat cosmology with 
$\Omega_M = 0.3$, $\Omega_\Lambda = 0.7$.  These transfer functions 
require a spatial scale parameter $\Gamma \sim \Omega_M h$ and a 
normalization which is commonly supplied by $\sigma_8$, the rms 
mass fluctuation between spherical volumes of radius $8\,h^{-1}\,$Mpc.
We find that typical shape parameters of $\Gamma = 0.16, 0.21, \hbox{and}\, 0.26$, and corresponding  normalizations of $\sigma_8 = 0.83, 0.94, 
\hbox{and}\, 1.05$ match these observations of bulk flows very well.

We find negligible difference whether we use a cube window function or
a top-hat, with side and radius related as above.  To demonstrate this
we show in Figure \ref{fig:bulky} open points which come from a
numerical simulation of a universe with the $\Gamma = 0.21$ and
$\sigma_8 = 0.94$ CDM power spectrum and random phases.  The bulk flows
in this simulation were computed in disjoint cubes precisely in the
same way as the data were treated (with the same assignment of top-hat
radii).  Since the simulated universe encompassed a volume which was
8192 times as large as the SBF survey volume, it was also possible to
assess the cosmic variance in bulk flow expected when velocity is
measured within the finite SBF survey volume.  This RMS variation, 
represented by the error bars on the simulation points in Figure
\ref{fig:bulky}, demonstrates that it will not be easy to put
stringent limits on the CDM transfer function shape parameter $\Gamma$
or normalization $\sigma_8$ individually, but that the combination
$\sigma_8 \Gamma^{-0.5}$ is fairly well constrained at $\sigma_8
\Gamma^{-0.5} \approx 2\pm0.5$.  This combination comes from the
constraint these data place on the slope of the power spectrum between
$k\sim0.2 h\,$Mpc$^{-1}$, which contributes to $\sigma_8$, and $k\sim0.04
h\,$Mpc$^{-1}$ which contributes to bulk flows.  The (one sigma) error
bar is dominated by the cosmic variance expected in the rather
small SBF survey volume.  The value of this measurement, of course, is
that it is directly sensitive to the mass fluctuations on scales of
50-200$h^{-1}$~Mpc without any recourse to galaxy distribution or bias
models.

In summary, the data of the SBF Survey seem completely compatible
with conventional models of the formation of large scale structure,
in contrast to the implications of even larger scale flows that
have been reported.

\section{Distance Independent Absolute Luminosities: $\overline N$}

Developing more accurate ways of measuring absolute luminosity for
galaxies that is independent of distance is obviously important for a
wide range of astronomical issues.  SBF is another, we believe, major
step in that program.  In the course of this study we have investigated
a further parameterization which is, in addition, independent of
photometric calibration or extinction as well.  This measure of the
absolute luminosity comes from the ratio of the total apparent flux
from the galaxy and the flux provided by the fluctuation signal.  We
express this in terms of magnitudes as the difference between the
fluctuation magnitude and the total magnitude of the galaxy:
\begin{equation}
  \overline N = \overline m - m_T.
  \label{eq:nbardef}
\end{equation}
We call \Nbar\ the ``fluctuation star count:'' it amounts to 
$+2.5\log_{10}$ of the total luminosity of the galaxy in units of the 
luminosity of a typical giant star.

We give values for \Nbar\ in the last columns of Tables 1 and 2.  The
total magnitudes were derived from the SBF survey photometry.  A
program written by B. Barris fitted a modified Sersic model (i.e.,
$\exp(-r^{1/n})$) and a sky offset to the azimuthally averaged profile
(after removal of stars and companion galaxies). The extrapolation to
infinity was converted to a total apparent magnitude.  The difference
between this total magnitude and the fluctuation magnitude yields
$\Nbar$.  As is well known, this extrapolation can be quite uncertain
because of the great extent of galaxies at low surface brightness;
probably for this reason there are occasionally significant 
discrepancies between our total magnitudes in the $I$ band and $B_T$ 
values from the RC3 (which generally are derived forcing $n$ to be 
$4$).  Therefore, we urge care in use of values of $I_T$ recovered 
from the values we give for \Nbar.

Not surprisingly, this absolute luminosity correlates with color;
Figure \ref{fig:nbarclr} shows the dependence of \vi\ on \Nbar.
\insfig{
\begin{figure}[t]
\epsscale{1.0}
\plotone{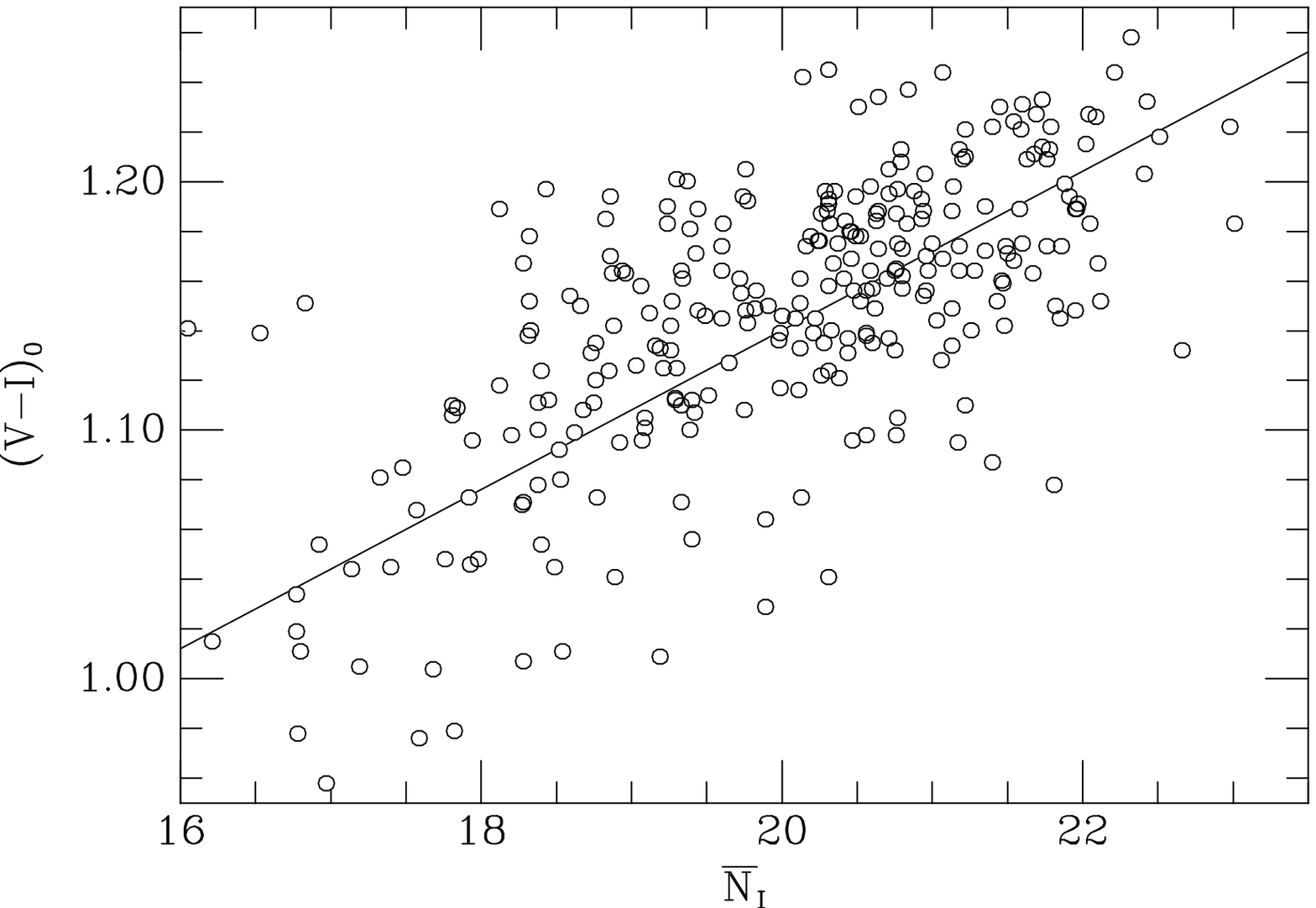}
\caption[nbarclr.eps]{
The \vi\ color of SBF galaxies is shown as a function of \Nbar.
\label{fig:nbarclr}}
\end{figure}
}
This correlation is approximated by 
\begin{equation}
  \vi = 0.50 + 0.032\Nbar
  \label{eq:nbarvi}
\end{equation}
The slope of the relation is so shallow that a very large error in
\Nbar\ has negligible effect on the prediction of \vi\ (a huge error of
0.5 mag in \Nbar\ corresponds to an error of only 0.016 mag in \vi).
The scatter in \vi\ for all galaxies is approximately 0.04 mag.  If we
restrict ourselves to only bona-fide elliptical galaxies with the best
measurements of \mi\ (error less than 0.30 mag) and \vi\ (error less
than 0.02 mag), the scatter decreases slightly to 0.03 mag.  Inasmuch
as the median error of the \vi\ values used is 0.015 mag, this means
that the intrinsic scatter in the prediction of \vi\ from \Nbar\ may be
as small as 0.025 mag.  This corresponds to an \RMS\ uncertainty of
only 0.02 mag in $E(B{-}V)$, which suggests that this is a promising
way to measure extinctions.

For example, by measuring the $I$ fluctuations in a galaxy which is
hidden behind 3 mag of extinction, one could establish
\Nbar\ independent of the extinction, derive \viz\ with an accuracy of
0.025 mag, \Mi\ with an uncertainty of 0.11 mag, $A_I$ with an accuracy
of 0.04 mag, and hence an SBF distance which has an error of only 0.12
mag from extinction.  Although there is some covariance in the final
answer from the use of \mi\ for both the intrinsic color and the
distance modulus, the slope of \vi\ with \Nbar\ is so shallow that the
covariance is very mild.

The direct use of \vi\ to estimate \Mi\ is operationally difficult
both because of the sensitivity to dust extinction and also the
necessity for very accurate photometry.  Use of \Nbar\ eases these 
requirements considerably.

In practice, this means that it is not necessary to go through \vi\ at all 
to estimate the absolute \Mi, although some color information or estimate 
of dust extinction is eventually necessary to get a distance modulus.
In order to demonstrate how \Mi\ depends on \Nbar\ we must once again
resort to groups so that \mi\ differs from  \Mi\ by only an offset.
Figure \ref{fig:nbar} shows values for \mi\ in eight groups as a function of
\Nbar. 
\insfig{
\begin{figure}[t]
\epsscale{1.0}
\plotone{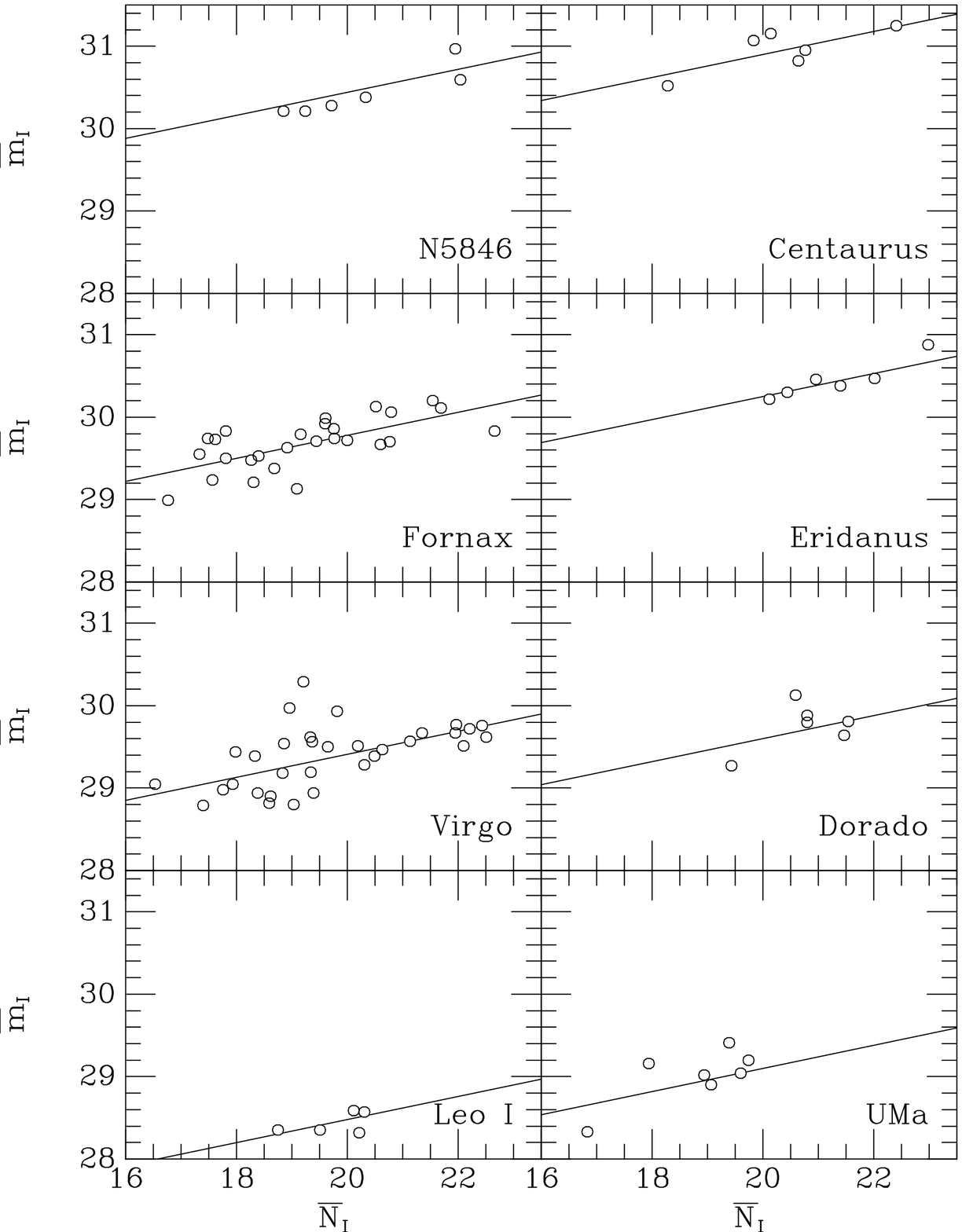}
\caption[nbar.eps]{
The fluctuation magnitude \mi\ for SBF galaxies in eight selected
groups is shown as a function of $\Nbar_I$.  The lines are {\it not}
fitted to the data, but are drawn with
a slope of 0.14 and zero points according to the SBF distance of the
group derived from \vi\ and \mi.
\label{fig:nbar}}
\end{figure}
}
The lines are all drawn using the SBF distance modulus to the group
based on the \vi\ measure of \Mi\ and the relation
\begin{equation}
  \Mi = -1.74 + 0.14(\Nbar-20).
  \label{eq:nbarmbar}
\end{equation}
Again, there is some covariance between \Mi\ derived this way and \mi,
but this correlation has such a shallow slope that a distance modulus
derived this way will suffer little in accuracy.

We do not want to suggest that use of \Nbar\ should supplant the use
of \vi\ to derive \Mi.  Use of \vi\ has a solid basis on stellar
populations whereas \Nbar\ correlates with \Mi\ for similar reasons that 
elliptical galaxies fall on a fundamental plane, and potentially has
unpleasant systematic problems.  Although the relation between \Mi\
and \Nbar\ shown in Figure \ref{fig:nbar} appears to track the
relation between  
\Mi\ and \vi\ in eight different groups, it would not be surprising if
the \Mi--\Nbar\ relation has environmental or type dependencies.  We
offer \Nbar\ mainly as a simple and easy way to get a fairly reliable
distance, and believe that it is a worthy subject for study in
its own right.  For example, Figure \ref{fig:nbarsig} illustrates the
rather tight correlation between \Nbar\ and central velocity
dispersion, tabulated by Prugniel and Simien (1996):
\insfig{
\begin{figure}[t]
\epsscale{1.0}
\plotone{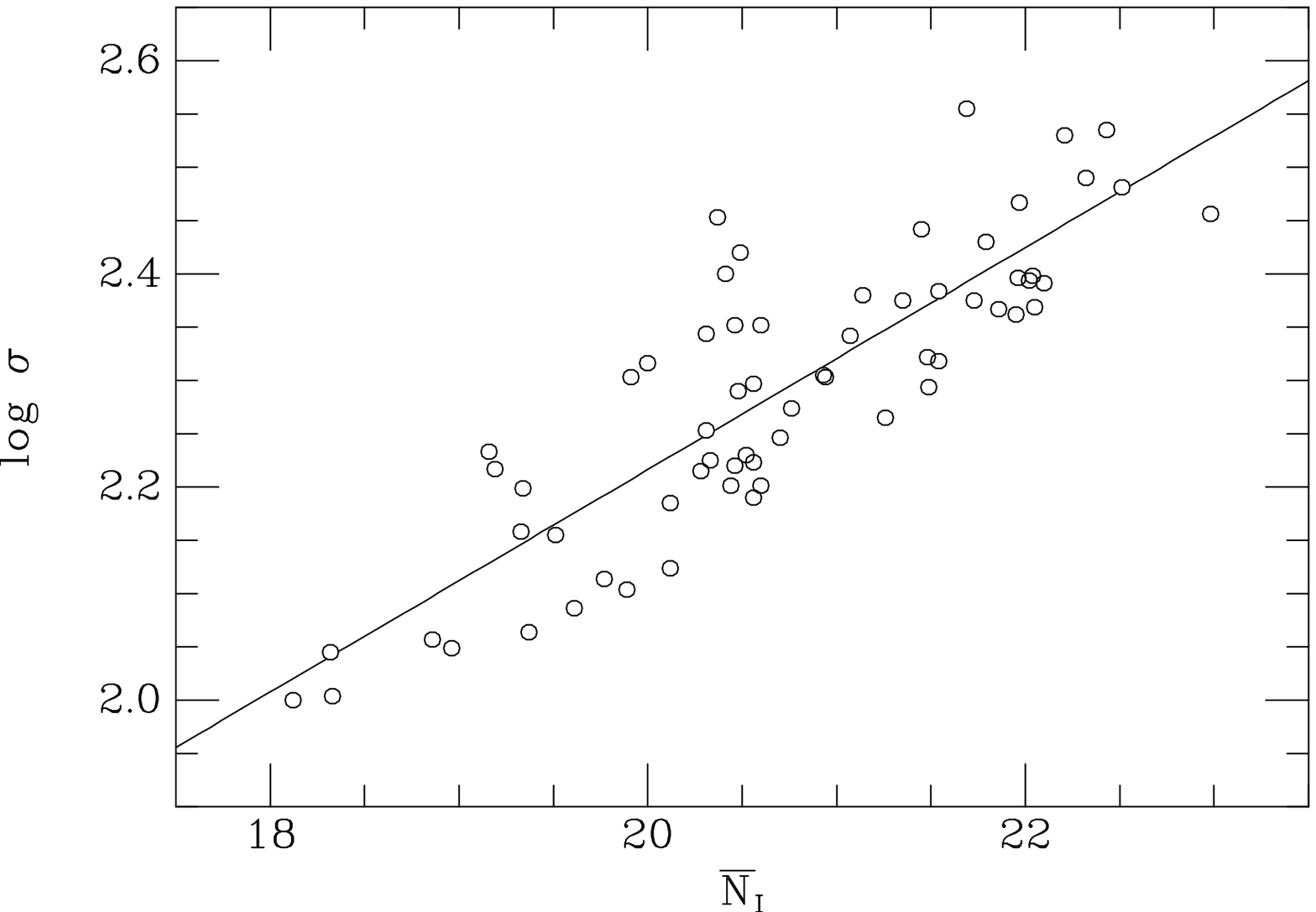}
\caption[nbarsig.eps]{
The velocity dispersion of elliptical SBF galaxies is shown as a
function of $\Nbar_I$.
\label{fig:nbarsig}}
\end{figure}
}
\begin{equation}
  \log\,\sigma = 2.22  + 0.10 \, (\Nbar - 20).
  \label{eq:nbarsig}
\end{equation}

\section{Lessons Learned and Future SBF Possibilities}

The SBF Survey was an unexpectedly large project.  We condensed a total 
of 7828 CCD frames taken over 65 observing runs to 2646 independent images,
comprising a total of 1022 hours of exposure time.  Unfortunately,
most of the integration time was spent on observations which
eventually did not contribute to measurements of \mi.  Either the
seeing was superseded by a later observation or the time was spent on
photometry, as discussed below.  The total observing time
that resulted in the \mi\ measurements of Table 1 amounted to only 370 
hours.  

Figure
\ref{fig:psfactual} shows the distribution of seeing during these
observations.
\insfig{
\begin{figure}[t]
\epsscale{1.0}
\plotone{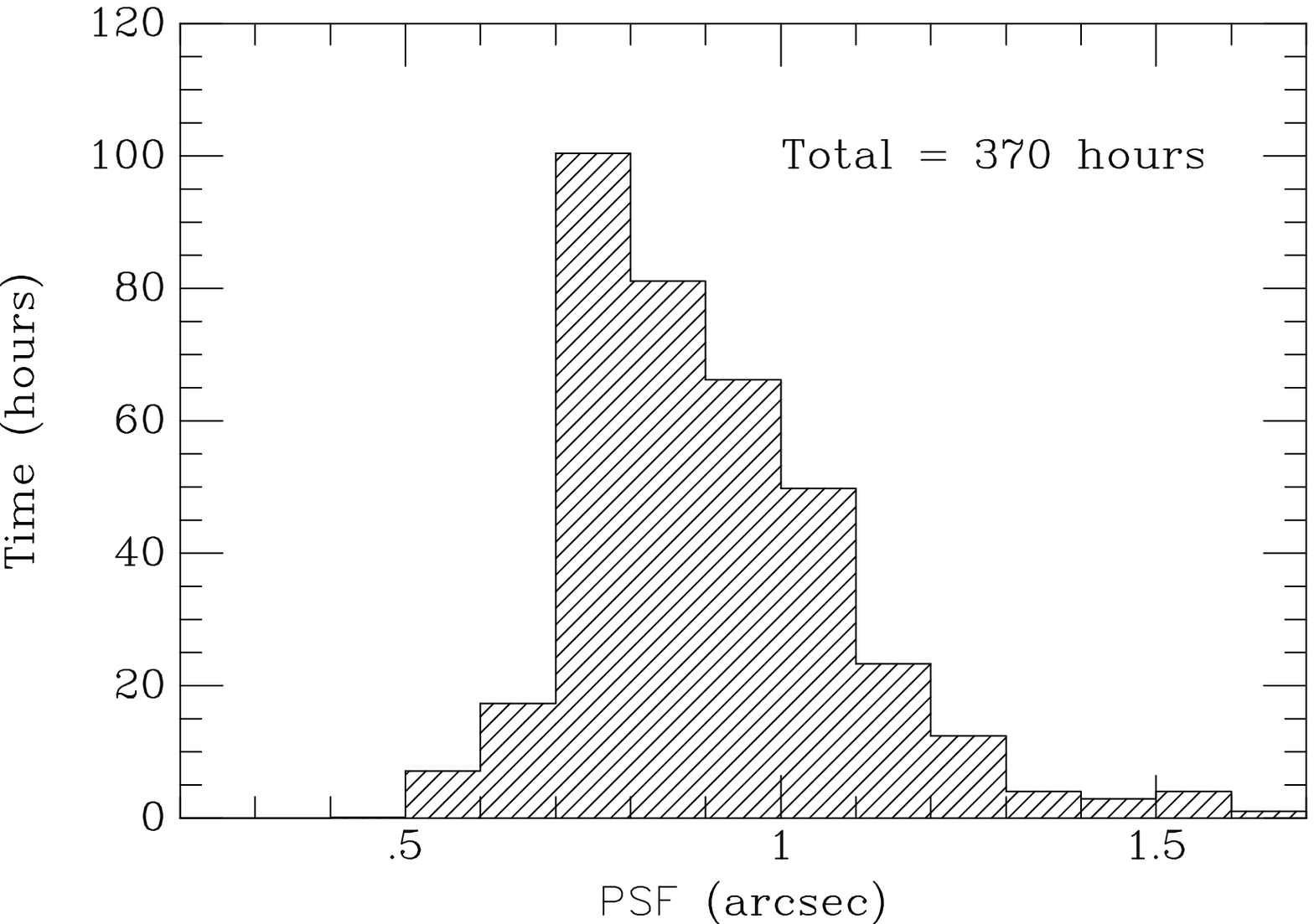}
\caption[psfactual.eps]{
The number of hours of exposure time is shown as a function of seeing
for the observations which eventually contributed to the tabulated
values for \mi.
\label{fig:psfactual}}
\end{figure}
}

Obviously, good seeing and high throughput are the keys to successful
SBF measurements.  If, for example, we were to take advantage of better
seeing to redo the present survey, reaching $\PD = 1.3$ by collecting
10 e$^-$ per \mi\ with a high-resistivity, near IR sensitive CCD and a
wider-bandpass, high throughput $I$ filter (gaining us about a factor
of two in sensitivity), we could redo the entire survey in about 100
hours on a 2.4-m telescope.  Furthermore, this set of would distance
measurements would be a factor of two better, with a median error in
\mi\ of $\sim$0.12 mag.

Figure 
\ref{fig:psfideal} shows the distribution of seeing we would require
for this project.
\insfig{
\begin{figure}[t]
\epsscale{1.0}
\plotone{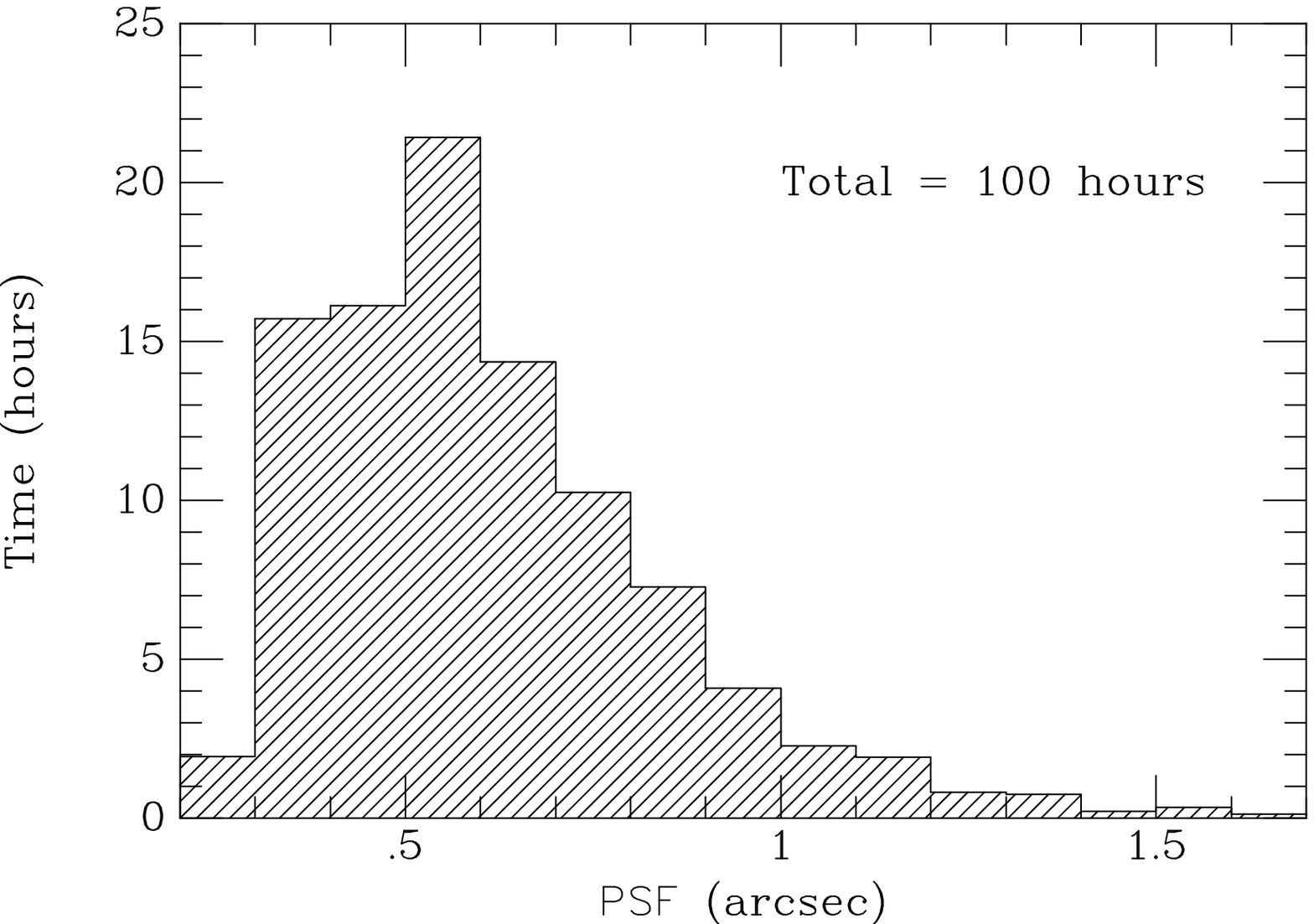}
\caption[psfideal.eps]{
The number of hours of exposure time is shown as a function of seeing
for a new set of observations which would improve the error in \mi\ to 
a median of 0.12 mag.
\label{fig:psfideal}}
\end{figure}
}

Another major lesson learned concerned photometry.  It is clear in
hindsight that I-band SBF is very color sensitive, since the K and M
giants that dominate the fluctuations are very red and have strong
variations in (V-I) color.  As a result, excellent photometry is
mandatory for SBF.  Our adopted program of ``fitting in'' photometric
observations into our \mi\ observations was not efficient.  It would
have been far better to devote all the good-seeing time on large
telescopes for \mi\ measurements and collect all the photometry on a few
photometric, well-calibrated nights with a 1-m class telescope, with a
common filter and CCD.  As discussed in SBF-I, we encountered a
particular problem calibrating I-band photometry with Tektronix CCDs
which Sirianni et al. (1998) suggest is due to reflections from the
mounting glass.  This results in a large halo around the psf which
affects points sources (standard stars) more than extended sources
(target galaxies).

Finally, better software could have helped.  We certainly learned how
best to carry out SBF reductions throughout the project, but software
which was more automated would have meant that improvements could be
easily applied to already reduced data.  Some aspects of the software
could stand improvement even today, particularly galaxy fitting.
Improved analytic models of systems with embedded disks, for example,
highly inclined S0 galaxies, would doubtless add more galaxies with
good SBF determinations, since removal of such large-scale features is
essential for a good measurement of the power spectrum on smaller
scales.  Our software is available on request to aid in the development
of more refined SBF reduction packages.

These lessons would have greatly eased and improved our ground-based
$I$-band SBF survey that we spent a decade completing, but we also
believe that the true potential of the SBF technique lies ahead with
new tools and applications.  Our successful SBF observations have
mostly been limited to galaxies with distances $D \lap 40$ Mpc.  With
the increased aperture size and better delivered image quality of the
new generation of large telescopes it should be possible to push
considerably further: in 0.4\arcsec\ seeing it will be straightforward
to get excellent measurements out beyond 50\,Mpc.  This is an
interesting range for measuring large scale flows, near enough that an
individual measurement has very good accuracy ($<200$\kms), yet far
enough to sample more than our local volume.  (We also note that the
accuracy of the distances we present here is limited mostly by
observing conditions; we believe it would even be worthwhile to redo
our survey given far superior observing conditions.  Figure
\ref{fig:psfideal} is similar to the seeing distribution hoped for on
the new generation of telescopes, and an aperture larger than 2.4-m
would reduce the necessary exposure times proportionately.

Recent improvements in size and characteristics in IR arrays have
opened the door for the SBF technique.  It is possible that the near-IR
will ultimately prove more powerful: the amplitude of the SBF signal is
20--40 times larger, the seeing is usually much better, and dust in the
target galaxy is less of a problem.  On the other hand, the much brighter 
sky detracts from these advantages, and there are concerns about 
variations in the stellar population, galaxy to galaxy, that may be more
difficult to calibrate.  Whereas the I-band is dominated by light from
every main sequence star as it climbs the giant branch, in the K-band
there is significant contribution from rarer AGB stars, which could
make distance measurements sensitive to stochastic variations in
intermediate-age populations. There is also substantial doubt as to
whether near-IR SBF magnitudes for elliptical galaxies are more than a
one-parameter family.  According to Blakeslee \etal\ (2000), the
stellar population models indicate that, at low metallicities, near-IR
SBF magnitudes depend mainly on age, while integrated color depend
mainly on metallicity. At higher metallicities, the SBF amplitudes
plateau, leading to a more complicated relation between color and
$\overline M$.   Depending upon the mixture of ages and metallicities
occurring in actual galaxies, this relation might or might not be a
simple one. Thus, an accurate characterization of near-IR SBF for a
large and diverse sample of galaxies could be extremely important from
the standpoint of both stellar population and distance studies.

The next generations of HST instruments (ACS, WFC3) should be superb
for SBF surveys.  The imaging quality of HST makes it possible to
get good $I$-band SBF distances at 10,000\kms. These new cameras will
be much more sensitive in the red than WFPC2, so the exposure times will
be shorter. The field of view will also be greater, giving more
surface area for the measurement.  Using the NICMOS camera SBF in
the $H$-band can reach 10,000\kms\ in a single orbit, and a modern IR
array in WFC3 would be even faster and better.

SBF has the potential to provide an accurate, unbiased chart of
large-scale structure --- defined by mass rather than galaxies ---
closer than 10,000\kms.  Apart from cosmological studies of $H_0$,
large scale flows, local inhomogeneities, biasing of galaxy density
with respect to mass, this will provide an important data for
understanding the physics of galaxy formation by relating galaxy
properties to the large-scale distribution of dark matter. No longer
limited by our inability to convert fluxes and angles to luminosities
and distances, we will add needed precision into studies of the history
of star formation and structure for the nearest galaxies, those that we
can study best.

\acknowledgements
It is with great sadness that we acknowledge the tragic death of our 
good friend and colleague Jeffrey Willick during the preparation of 
this paper.  Jeff's contributions to the study of large-scale flows
have been an important influence on our work; he will be sorely
missed.

We remain grateful to all our friends who have helped us collect
these data over the years.  Paul Schechter in particular has
contributed many observations and much advice.  Discussions with Nick
Kaiser about variance of variance were extremely helpful.  Thanks are
due to Brian Barris for providing us with profile fits and magnitudes.
This work was supported primarily by NSF grant AST9401519.

\clearpage
\setlength{\tabcolsep}{4pt}
\begin{deluxetable}{lrrrrrrrrrrrrrrr}
\tablewidth{0pt}
\tablecaption{SBF Data}
\scriptsize
\tablehead{
\colhead{Galaxy}   & \colhead{RA} & \colhead{Dec}    & \colhead{$v_{CMB}$} &
 \colhead{T} & \colhead{Grp}   & \colhead{$A_B$} & 
\colhead{$(V{-}I)$} & \colhead{$\overline m_I$}   & \colhead{$(m{-}M)$}   &
\colhead{$\langle r \rangle$} &
\colhead{$Q$}   & \colhead{$PD$} & \colhead{$\overline N_I$}
}
\startdata
N7814   &  0.813& 16.146&  684& 2&  0&0.19&1.245 0.017 2&29.29 0.10 2& 30.60 0.14&   36 0.4 2.4& 8.3&0.51&20.3\nl
N0063   &  4.440& 11.449&  803& 0&  0&0.48&0.979 0.018 2&28.85 0.31 1& 31.36 0.33&   24 0.6 1.3& 5.4&1.14&17.8\nl
N0147   &  8.298& 48.508& -456&-5&282&0.75&1.024 0.009 3&22.13 0.15 2& 24.44 0.16&   35 0.0 5.0& 9.9&0.01&13.6\nl
N0185   &  9.742& 48.338& -494&-5&282&0.79&1.051 0.017 4&21.83 0.13 3& 24.02 0.16&   43 0.0 4.1& 9.9&0.01&14.2\nl
N0221   & 10.675& 40.865& -494&-6&282&0.35&1.133 0.007 6&22.73 0.05 3& 24.55 0.08&   30 0.0 5.9& 9.9&0.01&15.6\nl
N0224   & 10.685& 41.269& -590& 3&282&0.35&1.231 0.007 5&23.03 0.05 3& 24.40 0.08&   51 0.0 3.4& 9.9&0.01&21.6\nl
N0274   & 12.758& -7.058& 1390&-3&  0&0.24&1.135 0.020 2&29.64 0.45 1& 31.45 0.47&   33 0.7 1.4& 3.0&1.68&18.8\nl
N0404   & 17.363& 35.718& -332&-3&  0&0.25&1.054 0.011 2&25.40 0.07 2& 27.57 0.10&   23 0.0 3.8& 9.9&0.01&16.9\nl
N0448   & 18.816& -1.625& 1589&-3&  0&0.26&1.132 0.029 1&30.58 0.33 1& 32.41 0.35&   18 0.4 1.6& 4.3&1.41&19.3\nl
N0524   & 21.199&  9.539& 2091&-1&  0&0.36&1.221 0.010 3&30.48 0.19 1& 31.90 0.20&   64 0.7 1.4& 2.6&1.65&21.6\nl
N0584   & 22.837& -6.868& 1566&-5& 26&0.18&1.157 0.009 4&29.82 0.19 2& 31.52 0.20&   23 0.3 3.4& 3.1&1.41&20.6\nl
N0596   & 23.217& -7.033& 1509&-4& 26&0.16&1.135 0.008 5&29.89 0.08 2& 31.69 0.10&   31 0.2 2.5& 3.1&1.51&20.3\nl
N0636   & 24.778& -7.513& 1504&-5& 26&0.11&1.156 0.008 5&30.65 0.14 1& 32.37 0.16&   19 0.4 3.0& 3.8&1.13&20.6\nl
N0720   & 28.252&-13.739& 1438&-5&  0&0.07&1.214 0.009 4&30.76 0.15 1& 32.21 0.17&   32 0.2 1.8& 4.0&1.09&21.7\nl
N0821   & 32.088& 10.996& 1433&-5&  0&0.47&1.196 0.022 1&30.38 0.13 1& 31.91 0.17&   30 0.5 2.2& 5.3&1.00&20.9\nl
N0855   & 33.515& 27.877&  338&-5&  0&0.31&1.015 0.018 1&27.59 0.14 2& 29.94 0.17&   19 0.5 1.8& 9.6&0.23&16.2\nl
N0891   & 35.638& 42.347&  305& 3&  0&0.28&1.142 0.017 1&27.83 0.11 1& 29.61 0.14&   94 0.5 1.9& 9.6&0.36&19.3\nl
N0936   & 36.907& -1.155& 1176&-1&  0&0.15&1.213 0.010 3&30.35 0.28 1& 31.81 0.28&   49 0.7 1.3& 5.4&0.94&21.8\nl
N0949   & 37.704& 37.136&  381& 3&  0&0.25&0.958 0.011 2&27.65 0.16 1& 30.26 0.18&   35 0.6 1.2& 9.3&0.44&17.0\nl
N1023   & 40.100& 39.063&  432&-3&  0&0.26&1.193 0.017 2&28.74 0.13 3& 30.29 0.16&   45 0.1 3.9& 9.5&0.25&20.9\nl
N1052   & 40.270& -8.256& 1242&-5&207&0.12&1.213 0.010 3&29.98 0.26 2& 31.44 0.27&   46 0.4 1.7& 3.5&1.48&20.8\nl
N1162   & 44.733&-12.399& 2126&-5& 29&0.21&1.173 0.032 1&31.44 0.27 1& 33.08 0.31&   18 0.4 3.3& 1.8&1.91&20.6\nl
N1172   & 45.400&-14.837& 1472&-4& 29&0.28&1.112 0.032 1&29.75 0.13 1& 31.66 0.20&   23 0.3 2.5& 3.8&1.40&19.4\nl
N1199   & 45.910&-15.614& 2512&-5& 29&0.23&1.188 0.012 3&31.03 0.32 1& 32.60 0.32&   28 0.3 2.1& 0.9&2.03&20.9\nl
N1201   & 46.035&-26.068& 1548&-2&  0&0.07&1.178 0.009 3&29.91 0.29 1& 31.53 0.30&   44 0.7 1.4& 1.9&2.04&20.5\nl
N1209   & 46.513&-15.612& 2429&-5& 29&0.16&1.198 0.010 3&31.24 0.22 1& 32.77 0.23&   20 0.4 2.9& 1.7&1.72&21.1\nl
N1297   & 49.809&-19.101& 1383&-2&  0&0.12&1.187 0.018 2&30.70 0.41 1& 32.28 0.42&   44 0.7 1.4& 3.4&1.42&20.3\nl
N1316   & 50.673&-37.208& 1657&-2& 31&0.09&1.132 0.016 2&29.83 0.15 2& 31.66 0.17&  110 0.3 2.2& 1.9&1.79&22.7\nl
I1919   & 51.508&-32.896& 1058&-3& 31&0.06&1.108 0.018 1&29.38 0.14 1& 31.31 0.17&   16 0.4 1.8& 5.5&0.85&18.7\nl
N1332   & 51.572&-21.336& 1316&-3& 32&0.14&1.222 0.010 4&30.38 0.16 1& 31.80 0.18&   37 0.4 1.6& 4.1&1.01&21.4\nl
N1336   & 51.630&-35.714& 1344&-3& 31&0.05&1.124 0.032 1&29.53 0.15 2& 31.38 0.21&   34 0.2 1.7& 3.7&1.45&18.4\nl
E358-006& 51.824&-34.527& 1234&-4& 31&0.04&1.068 0.020 2&29.24 0.31 1& 31.35 0.32&   13 0.6 2.2& 4.8&0.97&17.6\nl
N1339   & 52.027&-32.286& 1252&-4& 31&0.06&1.134 0.012 3&29.79 0.35 1& 31.61 0.35&   20 0.4 1.5& 4.2&0.95&19.2\nl
N1344   & 52.080&-31.068& 1086&-5& 31&0.08&1.135 0.011 3&29.67 0.29 2& 31.48 0.30&   27 0.3 2.2& 5.0&0.97&20.6\nl
N1351   & 52.645&-34.853& 1420&-3& 31&0.06&1.148 0.016 2&29.86 0.13 1& 31.61 0.16&   26 0.3 2.2& 3.8&1.12&19.8\nl
N1366   & 53.472&-31.193& 1186&-2& 31&0.07&1.095 0.018 1&29.63 0.27 1& 31.62 0.29&   22 0.7 1.4& 3.6&1.48&18.9\nl
N1373   & 53.745&-35.171& 1272&-4& 31&0.06&1.085 0.013 2&29.74 0.46 1& 31.78 0.47&   11 0.7 1.4& 2.7&2.05&17.5\nl
N1375   & 53.819&-35.266&  660&-2& 31&0.06&1.070 0.019 1&29.48 0.09 3& 31.58 0.13&   23 0.3 2.5& 8.2&0.51&18.3\nl
N1374   & 53.820&-35.226& 1284&-5& 31&0.06&1.146 0.016 2&29.72 0.10 1& 31.48 0.13&   37 0.2 3.1& 4.3&1.00&20.0\nl
E358-025& 53.889&-32.465& 1346&-3& 31&0.04&1.034 0.018 1&28.99 0.24 1& 31.25 0.26&   23 0.7 1.3& 2.4&2.07&16.8\nl
N1379   & 54.014&-35.441& 1297&-5& 31&0.05&1.143 0.019 1&29.74 0.11 1& 31.51 0.15&   41 0.2 2.8& 4.2&1.00&19.8\nl
N1380   & 54.112&-34.976& 1732&-2& 31&0.08&1.197 0.019 1&29.70 0.15 2& 31.23 0.18&   49 0.3 2.4& 0.8&2.67&20.8\nl
N1381   & 54.131&-35.294& 1640&-2& 31&0.06&1.189 0.018 1&29.71 0.19 1& 31.28 0.21&   24 0.6 1.2& 3.1&1.16&19.4\nl
N1386   & 54.193&-35.999&  815&-1& 31&0.05&1.101 0.018 1&29.13 0.23 1& 31.09 0.25&   46 0.6 1.3& 5.3&1.26&19.1\nl
N1380A  & 54.197&-34.739& 1485&-2& 31&0.06&1.138 0.018 1&29.21 0.28 1& 31.00 0.29&   23 0.6 1.3& 2.1&2.15&18.3\nl
N1387   & 54.238&-35.506& 1192&-3& 31&0.06&1.208 0.047 2&30.06 0.14 1& 31.54 0.26&   26 0.3 2.2& 4.3&1.16&20.8\nl
N1382   & 54.285&-35.195& 1652&-3& 31&0.07&1.106 0.013 2&29.83 0.30 2& 31.77 0.31&   18 0.4 2.1& 1.2&2.58&17.8\nl
N1389   & 54.299&-35.745&  841&-3& 31&0.05&1.145 0.019 1&29.92 0.16 1& 31.68 0.18&   26 0.6 2.4& 6.1&0.91&19.6\nl
N1399   & 54.621&-35.449& 1314&-5& 31&0.06&1.227 0.016 2&30.11 0.13 1& 31.50 0.16&   46 0.2 2.6& 3.8&1.34&21.7\nl
N1395   & 54.623&-23.028& 1568&-5& 32&0.10&1.215 0.010 4&30.47 0.14 2& 31.91 0.16&   40 0.2 2.9& 2.9&1.49&22.0\nl
N1411   & 54.688&-44.100&  910&-3&  0&0.04&1.107 0.018 1&29.54 0.23 1& 31.48 0.25&   30 0.5 2.0& 4.9&1.21&19.4\nl
N1404   & 54.715&-35.593& 1817&-5& 31&0.05&1.224 0.016 2&30.20 0.16 1& 31.61 0.19&   45 0.3 2.7& 2.0&2.00&21.5\nl
N1400   & 54.880&-18.689&  409&-3& 32&0.28&1.170 0.009 4&30.46 0.32 3& 32.11 0.33&   34 0.4 2.3& 9.6&0.32&21.0\nl
N1407   & 55.052&-18.581& 1627&-5& 32&0.30&1.222 0.010 4&30.88 0.25 1& 32.30 0.26&   31 0.2 1.9& 2.9&1.42&23.0\nl
N1419   & 55.178&-37.512& 1456&-5& 31&0.06&1.110 0.018 1&29.50 0.22 1& 31.42 0.24&   22 0.7 1.4& 2.8&1.63&17.8\nl
N1427   & 55.582&-35.393& 1326&-4& 31&0.05&1.152 0.018 1&30.13 0.22 1& 31.86 0.24&   42 0.4 1.4& 3.5&1.45&20.5\nl
N1426   & 55.705&-22.110& 1313&-5& 32&0.07&1.161 0.009 4&30.22 0.17 1& 31.91 0.18&   29 0.2 2.0& 4.3&1.27&20.1\nl
N1439   & 56.210&-21.922& 1543&-5& 32&0.13&1.131 0.009 4&30.30 0.13 1& 32.13 0.15&   28 0.3 2.1& 3.3&1.25&20.4\nl
E358-059& 56.264&-35.973&  931&-3& 31&0.04&1.081 0.016 2&29.55 0.18 1& 31.60 0.20&   13 0.6 2.3& 6.4&1.02&17.3\nl
I2006   & 58.618&-35.967& 1252&-5& 31&0.05&1.183 0.018 1&29.99 0.27 1& 31.59 0.29&   45 0.7 1.4& 3.6&1.45&19.6\nl
N1527   & 62.102&-47.897& 1117&-3&  0&0.05&1.230 0.016 2&29.90 0.20 1& 31.28 0.22&   25 0.3 2.3& 4.4&0.95&20.5\nl
N1533   & 62.464&-56.121&  744&-3&211&0.07&1.198 0.018 1&30.13 0.37 1& 31.65 0.38&   44 0.7 1.4& 5.7&1.14&20.6\nl
N1543   & 63.180&-57.737& 1065&-2&211&0.12&1.173 0.016 2&29.88 0.14 1& 31.51 0.17&   52 0.3 2.2& 4.2&1.11&20.8\nl
N1537   & 63.421&-31.646& 1300&-3&210&0.11&1.096 0.018 1&29.84 0.16 1& 31.83 0.19&   40 0.4 3.0& 3.0&1.72&20.5\nl
N1549   & 63.938&-55.592& 1128&-5&211&0.05&1.168 0.016 2&29.81 0.15 1& 31.47 0.18&   24 0.3 2.4& 3.5&1.13&21.5\nl
N1553   & 64.043&-55.781& 1256&-2&211&0.05&1.159 0.016 2&29.64 0.14 1& 31.34 0.17&   29 0.2 2.0& 3.3&1.31&21.5\nl
N1574   & 65.496&-56.974& 1025&-3&211&0.07&1.162 0.016 2&29.80 0.19 1& 31.49 0.21&   35 0.4 1.7& 4.0&1.10&20.8\nl
N1596   & 66.907&-55.027& 1508&-2&211&0.04&1.171 0.016 2&29.27 0.13 1& 30.92 0.16&   21 0.3 2.8& 2.2&1.57&19.4\nl
N2271   &100.718&-23.476& 2724&-3&  0&0.53&1.234 0.029 1&31.15 0.25 1& 32.51 0.28&   21 0.4 2.8& 1.9&1.93&20.6\nl
N2293   &101.929&-26.753& 2135&-1&  0&0.52&1.237 0.013 2&29.81 0.40 1& 31.16 0.41&   11 0.7 1.4& 2.6&2.03&20.8\nl
N2325   &105.669&-28.697& 2408&-5&  0&0.51&1.164 0.014 2&30.36 0.14 1& 32.04 0.16&   19 0.3 2.3& 2.1&2.14&21.3\nl
N2380   &110.980&-27.529& 1963&-2&  0&1.32&1.110 0.009 4&30.14 0.35 1& 32.05 0.35&   15 0.5 1.9& 2.1&1.63&21.2\nl
E208-021&113.486&-50.442& 1166&-3&  0&0.75&1.146 0.018 1&28.90 0.18 1& 30.66 0.20&   41 0.7 1.5& 4.0&1.43&19.5\nl
N2434   &113.715&-69.284& 1440&-5&212&1.07&1.098 0.055 1&29.70 0.15 1& 31.67 0.29&   27 0.3 2.2& 4.1&1.43&20.6\nl
I2311   &124.690&-25.371& 2090&-5&  0&0.62&1.136 0.027 1&30.04 0.14 1& 31.84 0.20&   19 0.4 3.0& 2.3&2.09&20.0\nl
N2549   &124.743& 57.803& 1159&-2&  0&0.28&1.163 0.012 2&28.82 0.27 1& 30.51 0.28&   33 0.7 1.3& 3.0&1.39&18.9\nl
N2592   &126.783& 25.971& 2213&-5&  0&0.26&1.205 0.010 3&30.55 0.42 2& 32.05 0.43&   28 0.4 1.6& 1.1&2.26&19.8\nl
N2634   &132.105& 73.967& 2309&-5&283&0.09&1.154 0.022 1&30.90 0.67 1& 32.62 0.68&   24 0.7 1.4& 1.5&2.22&20.9\nl
N2683   &133.171& 33.417&  640& 3&  0&0.14&1.147 0.015 1&27.68 0.35 1& 29.44 0.36&   48 0.7 1.5& 5.4&0.49&19.1\nl
N2681   &133.388& 51.315&  868& 0&  0&0.10&1.041 0.010 3&28.95 0.34 1& 31.18 0.34&   60 0.6 2.4& 6.1&0.74&20.3\nl
N2695   &133.613& -3.067& 2126&-2&273&0.08&1.183 0.051 1&30.96 0.32 2& 32.55 0.40&   20 0.3 2.9& 3.1&1.64&20.3\nl
N2699   &133.953& -3.128& 2128&-5&273&0.09&1.152 0.051 1&30.42 0.14 1& 32.15 0.27&   18 0.4 1.6& 2.2&1.68&19.3\nl
N2768   &137.907& 60.039& 1483&-5&215&0.19&1.144 0.027 1&29.98 0.20 1& 31.75 0.24&   24 0.7 1.4& 3.4&0.96&21.0\nl
N2784   &138.078&-24.173& 1014&-2&  0&0.93&1.188 0.023 1&28.40 0.23 2& 29.96 0.25&   54 0.3 2.2& 4.4&1.37&20.3\nl
N2778   &138.102& 35.028& 2250&-5&216&0.09&1.150 0.015 1&30.06 0.29 1& 31.80 0.30&   18 0.5 2.0& 1.5&2.07&18.7\nl
N2787   &139.829& 69.203&  763&-1&  0&0.57&1.194 0.019 1&27.82 0.35 1& 29.37 0.36&   35 0.6 1.3& 3.3&1.18&18.9\nl
N2865   &140.878&-23.163& 2897&-5&284&0.36&1.105 0.019 2&30.95 0.17 1& 32.89 0.20&   21 0.3 2.8& 0.6&2.14&20.8\nl
N2880   &142.396& 62.490& 1677&-3&215&0.14&1.148 0.015 1&29.95 0.20 1& 31.70 0.21&   22 0.4 1.6& 2.9&1.31&19.4\nl
N2904   &142.570&-30.384& 2694&-3&  0&0.54&1.201 0.041 1&30.36 0.14 2& 31.86 0.24&   19 0.3 2.2& 1.7&2.07&19.3\nl
N2974   &145.639& -3.700& 2266&-5&  0&0.23&1.203 0.015 1&30.16 0.23 1& 31.66 0.24&   37 0.6 2.4& 0.6&2.15&21.0\nl
N2950   &145.652& 58.852& 1466&-2&  0&0.07&1.110 0.019 1&28.95 0.25 1& 30.87 0.27&   33 0.7 1.3& 1.0&2.23&19.3\nl
N3032   &148.033& 29.237& 1842&-2&  0&0.07&1.073 0.019 1&29.62 0.26 1& 31.71 0.28&   25 0.7 1.4& 2.7&1.71&18.8\nl
N3056   &148.637&-28.297& 1373&-1&  0&0.39&1.073 0.023 1&28.34 0.22 1& 30.43 0.25&   27 0.6 2.2& 2.5&1.98&17.9\nl
N3031   &148.890& 69.067&   46& 2&  0&0.35&1.187 0.011 3&26.38 0.25 3& 27.96 0.26&   72 0.0 2.4& 9.9&0.04&20.8\nl
N3078   &149.602&-26.926& 2837&-5&219&0.31&1.209 0.017 2&31.25 0.29 2& 32.73 0.30&   31 0.5 2.0& 1.0&2.67&21.6\nl
N3087   &149.787&-34.225& 2976&-4&218&0.45&1.164 0.019 2&31.04 0.18 1& 32.72 0.21&   23 0.3 2.6& 1.6&2.23&21.2\nl
N3073   &150.216& 55.620& 1317&-3&  0&0.04&1.007 0.019 1&30.26 0.92 1& 32.64 0.93&   17 0.7 1.3& 1.9&1.51&18.3\nl
N3077   &150.838& 68.734&   95& 0&  0&0.29&1.044 0.015 2&25.81 0.10 1& 28.03 0.13&   46 0.4 2.9& 9.9&0.09&17.1\nl
N3115   &151.309& -7.719& 1054&-3&  0&0.20&1.183 0.010 6&28.34 0.06 4& 29.93 0.09&   54 0.3 3.2& 4.4&0.95&20.8\nl
N3136   &151.451&-67.378& 1823&-5& 44&1.03&1.095 0.033 2&29.96 0.15 1& 31.95 0.22&   32 0.5 1.9& 3.5&1.39&21.2\nl
N3136B  &152.555&-67.005& 1949&-4& 44&0.81&1.164 0.033 2&29.96 0.11 2& 31.64 0.19&   17 0.4 3.4& 3.1&1.52&19.6\nl
N3156   &153.171&  3.131& 1651&-2&252&0.15&1.011 0.011 5&29.38 0.13 1& 31.75 0.14&   17 0.4 1.7& 3.0&1.34&18.5\nl
N3193   &154.604& 21.895& 1696&-5& 45&0.11&1.174 0.009 4&31.03 0.17 1& 32.66 0.18&   23 0.2 3.0& 3.8&0.85&21.5\nl
N3226   &155.864& 19.899& 1601&-5& 45&0.10&1.178 0.015 1&30.25 0.23 1& 31.86 0.24&   23 0.5 1.9& 2.7&1.33&20.5\nl
N3250   &156.635&-39.943& 3189&-5& 46&0.44&1.226 0.019 3&31.75 0.14 2& 33.15 0.17&   27 0.3 2.2& 1.6&2.14&22.1\nl
N3245   &156.826& 28.508& 1657&-2&  0&0.11&1.139 0.023 1&29.81 0.16 1& 31.60 0.20&   25 0.4 1.7& 1.2&1.79&20.0\nl
N3257   &157.196&-35.658& 3344&-3& 46&0.33&1.192 0.041 1&31.13 0.18 1& 32.68 0.26&   15 0.5 2.0& 1.2&2.61&19.8\nl
N3258   &157.226&-35.606& 3129&-5& 46&0.36&1.209 0.034 1&31.06 0.22 1& 32.53 0.27&   26 0.3 2.3& 1.7&2.32&21.2\nl
N3268   &157.503&-35.325& 3084&-5& 46&0.45&1.189 0.023 2&31.14 0.22 1& 32.71 0.25&    9 0.5 1.9& 1.7&1.85&21.6\nl
N3318   &159.315&-41.628& 3071& 3&  0&0.34&1.073 0.101 1&30.61 0.19 1& 32.70 0.49&   18 0.3 2.3& 1.1&2.06&20.1\nl
N3368   &161.688& 11.821& 1252& 2& 57&0.11&1.145 0.015 1&28.32 0.20 1& 30.08 0.22&  132 0.7 1.3& 1.7&1.63&20.2\nl
N3377   &161.924& 13.983& 1038&-5& 57&0.15&1.114 0.009 4&28.35 0.06 3& 30.25 0.09&   42 0.2 3.7& 4.8&0.89&19.5\nl
N3379   &161.958& 12.582& 1274&-5& 57&0.10&1.193 0.015 1&28.57 0.07 2& 30.12 0.11&   57 0.2 3.1& 2.4&1.59&20.3\nl
N3384   &162.072& 12.630& 1080&-3& 57&0.11&1.151 0.018 1&28.59 0.10 1& 30.32 0.14&   63 0.3 2.4& 3.9&1.24&20.1\nl
N3412   &162.722& 13.413& 1218&-2& 57&0.12&1.111 0.015 1&28.35 0.11 1& 30.27 0.14&   40 0.6 2.2& 3.0&1.40&18.7\nl
N3414   &162.818& 27.976& 1784&-2&  0&0.10&1.149 0.019 1&30.27 0.32 1& 32.01 0.33&   32 0.7 1.4& 1.2&1.66&20.6\nl
N3457   &163.703& 17.622& 1497& 0&  0&0.14&1.098 0.015 1&29.60 0.13 2& 31.58 0.15&   17 0.5 2.1& 3.4&1.32&18.2\nl
N3489   &165.076& 13.902& 1045&-1&  0&0.07&1.041 0.023 1&28.18 0.10 2& 30.41 0.15&   49 0.4 1.4& 3.9&0.96&18.9\nl
N3557   &167.490&-37.538& 3337&-5&154&0.43&1.183 0.016 1&31.72 0.20 1& 33.30 0.22&   20 0.3 2.1& 1.7&2.67&23.0\nl
N3585   &168.320&-26.756& 1845&-5&285&0.28&1.160 0.016 2&29.81 0.16 1& 31.51 0.18&   56 0.5 2.1& 1.2&2.31&21.5\nl
N3599   &168.864& 18.113& 1192&-2& 48&0.09&1.112 0.012 2&29.63 0.16 1& 31.54 0.18&   27 0.4 1.6& 4.4&1.14&19.3\nl
N3605   &169.195& 18.018& 1029&-5& 48&0.09&1.118 0.024 2&29.70 0.27 1& 31.58 0.29&    8 0.7 1.4& 5.1&1.11&18.1\nl
N3607   &169.225& 18.053& 1294&-2& 48&0.09&1.152 0.010 3&30.06 0.15 1& 31.79 0.17&   32 0.3 2.7& 3.8&1.40&21.4\nl
N3608   &169.245& 18.149& 1539&-5& 48&0.09&1.156 0.009 5&30.09 0.12 1& 31.80 0.14&   25 0.2 2.7& 3.8&0.92&20.5\nl
N3610   &169.608& 58.787& 1922&-5&281&0.04&1.108 0.015 1&29.73 0.20 1& 31.65 0.22&   32 0.7 1.4& 0.7&2.21&19.8\nl
N3613   &169.651& 58.001& 2216&-5&281&0.05&1.175 0.015 1&30.69 0.39 1& 32.32 0.40&   27 0.4 1.6& 0.3&2.22&21.0\nl
N3626   &170.015& 18.358& 1815&-1& 48&0.09&1.009 0.015 1&29.13 0.23 1& 31.51 0.24&   52 0.7 1.4& 1.6&2.03&19.2\nl
N3640   &170.278&  3.236& 1674&-5& 50&0.19&1.140 0.009 5&30.37 0.12 1& 32.16 0.13&   25 0.3 2.3& 3.1&1.49&21.3\nl
N3641   &170.286&  3.195& 2130&-5& 50&0.18&1.131 0.012 3&30.30 0.24 1& 32.13 0.25&   16 0.4 1.8& 1.8&1.90&18.7\nl
N3818   &175.489& -6.156& 1874&-5&153&0.16&1.124 0.015 1&30.94 0.60 1& 32.80 0.61&   32 0.7 1.4& 2.5&1.69&20.3\nl
N3904   &177.305&-29.276& 2095&-5& 52&0.31&1.156 0.055 1&30.55 0.13 1& 32.26 0.28&   26 0.3 2.2& 2.7&1.51&21.0\nl
N3923   &177.759&-28.806& 1953&-5& 52&0.36&1.194 0.055 1&30.26 0.12 1& 31.80 0.28&   42 0.2 2.8& 2.7&1.50&21.9\nl
N3928   &177.946& 48.681& 1187& 3&155&0.09&1.096 0.015 1&29.16 0.63 1& 31.14 0.64&   20 0.6 2.2& 2.5&1.47&17.9\nl
N3941   &178.230& 36.987& 1215&-2&  0&0.09&1.125 0.013 2&28.58 0.16 1& 30.43 0.18&   48 0.1 1.9& 1.3&1.63&19.3\nl
I0745   &178.551&  0.136& 1506&-2&  0&0.09&0.978 0.010 4&28.80 0.30 1& 31.31 0.30&   17 0.6 1.3& 0.2&2.33&16.8\nl
N3990   &179.401& 55.459&  879&-3&155&0.07&1.151 0.019 1&28.33 0.27 1& 30.06 0.28&   11 0.5 2.0& 4.2&0.98&16.8\nl
N3998   &179.486& 55.454& 1202&-2&155&0.07&1.194 0.011 3&29.20 0.18 1& 30.75 0.19&   34 0.3 2.6& 2.4&1.35&19.7\nl
N4026   &179.857& 50.962& 1077&-2&155&0.09&1.174 0.015 1&29.04 0.26 1& 30.67 0.28&   26 0.7 1.3& 4.7&0.86&19.6\nl
N4033   &180.145&-17.842& 1886&-5& 53&0.20&1.113 0.023 1&29.71 0.20 1& 31.62 0.23&   22 0.7 1.4& 1.2&2.56&19.3\nl
N4105   &181.670&-29.762& 2222&-5& 49&0.26&1.171 0.017 2&30.47 0.14 1& 32.12 0.17&   34 0.2 1.8& 2.4&1.89&21.5\nl
N4111   &181.761& 43.067& 1045&-1&155&0.06&1.096 0.015 1&28.90 0.22 1& 30.88 0.23&   34 0.6 1.3& 3.9&1.20&19.1\nl
N4125   &182.030& 65.173& 1455&-5& 54&0.08&1.174 0.011 2&30.26 0.24 1& 31.89 0.25&   52 0.2 1.7& 2.3&1.86&21.9\nl
N4138   &182.378& 43.688& 1070&-1&155&0.06&1.164 0.013 2&29.02 0.25 1& 30.70 0.26&   37 0.6 1.2& 3.0&1.01&18.9\nl
N4143   &182.402& 42.536& 1207&-2&155&0.05&1.181 0.015 1&29.41 0.17 1& 31.01 0.19&   24 0.4 1.8& 3.0&1.38&19.4\nl
N4150   &182.640& 30.402&  538&-2& 55&0.08&1.071 0.017 1&28.60 0.22 1& 30.69 0.24&   33 0.7 1.3& 6.3&0.81&18.3\nl
N4203   &183.772& 33.198& 1399&-3&  0&0.05&1.195 0.019 1&29.36 0.15 1& 30.90 0.18&   39 0.3 2.2& 1.9&1.26&20.7\nl
N4251   &184.533& 28.176& 1367&-2& 55&0.10&1.117 0.015 1&29.57 0.18 1& 31.46 0.20&   35 0.3 2.5& 2.9&1.41&20.0\nl
N4258   &184.741& 47.304&  664& 4&  0&0.07&1.134 0.023 1&27.50 0.08 2& 29.31 0.14&  103 0.7 1.7& 6.3&0.61&21.1\nl
N4261   &184.845&  5.827& 2557&-5&150&0.08&1.258 0.014 2&31.25 0.18 1& 32.50 0.19&   25 0.2 2.7& 1.0&1.28&22.3\nl
N4278   &185.030& 29.280&  938&-5& 55&0.12&1.161 0.012 4&29.34 0.18 1& 31.03 0.20&   40 0.4 1.8& 5.9&0.66&20.4\nl
N4291   &185.076& 75.373& 1765&-5& 98&0.16&1.175 0.017 1&30.47 0.31 1& 32.09 0.32&   33 0.3 2.1& 2.2&1.68&20.4\nl
N4283   &185.087& 29.311& 1371&-5& 55&0.11&1.178 0.010 3&29.37 0.18 1& 30.98 0.19&   18 0.6 2.2& 2.4&1.65&18.3\nl
N4346   &185.867& 46.994&  977&-2&  0&0.06&1.158 0.012 2&29.07 0.15 1& 30.78 0.17&   31 0.7 1.4& 4.0&1.00&19.1\nl
N4339   &185.893&  6.082& 1653&-5& 56&0.11&1.200 0.015 1&29.56 0.16 1& 31.08 0.18&   21 0.5 2.1& 1.1&2.07&19.4\nl
N4365   &186.116&  7.318& 1592&-5&150&0.09&1.222 0.017 1&30.14 0.14 1& 31.55 0.17&   35 0.3 2.5& 3.0&1.35&21.8\nl
N4386   &186.122& 75.530& 1698&-2&  0&0.17&1.196 0.019 1&30.63 0.48 1& 32.16 0.49&   34 0.6 1.3& 2.4&1.87&20.3\nl
N4374   &186.265& 12.887& 1375&-5& 56&0.17&1.191 0.008 5&29.77 0.09 2& 31.32 0.11&   55 0.2 2.8& 4.1&1.30&22.0\nl
N4379   &186.312& 15.608& 1407&-3& 56&0.10&1.185 0.017 1&29.18 0.39 1& 30.76 0.41&   28 0.4 1.6& 2.4&1.42&18.8\nl
N4391   &186.329& 64.934& 1451&-3& 54&0.08&1.112 0.015 1&29.99 0.26 1& 31.90 0.27&   16 0.7 1.4& 3.2&1.38&18.5\nl
N4382   &186.353& 18.191& 1087&-1&  0&0.13&1.150 0.022 3&29.59 0.09 2& 31.33 0.14&   53 0.1 2.7& 5.5&0.76&21.8\nl
E322-008&186.407&-39.320& 3335&-2& 35&0.36&1.205 0.023 2&31.37 0.50 1& 32.86 0.52&   21 0.7 1.4& 0.0&2.53&20.7\nl
N4387   &186.424& 12.812&  925&-5& 56&0.14&1.163 0.011 2&29.97 0.72 1& 31.65 0.73&   15 0.4 1.4& 3.0&1.46&19.0\nl
N4406   &186.549& 12.947&  120&-5& 56&0.13&1.167 0.008 5&29.51 0.12 1& 31.17 0.14&   27 0.2 2.8& 5.7&0.85&22.1\nl
N4419   &186.737& 15.048&  154& 1& 56&0.14&1.126 0.026 2&28.80 0.21 1& 30.65 0.25&   49 0.4 1.4& 2.5&1.15&19.0\nl
N4441   &186.837& 64.800& 1553&-1&  0&0.09&1.005 0.012 2&29.00 0.44 1& 31.40 0.45&   32 0.7 1.4& 1.5&1.51&17.2\nl
N4434   &186.904&  8.155& 1418&-5& 56&0.10&1.125 0.015 1&30.29 0.15 1& 32.14 0.17&   12 0.6 2.2& 6.9&0.61&19.2\nl
I3370   &186.904&-39.338& 3276&-5& 35&0.40&1.185 0.022 3&30.56 0.14 1& 32.14 0.17&   19 0.3 2.2& 0.9&2.69&20.9\nl
N4460   &187.192& 44.863&  782&-1&  0&0.08&1.011 0.015 1&27.54 0.17 1& 29.91 0.19&   33 0.7 1.4& 5.6&0.88&16.8\nl
N4458   &187.241& 13.243& 1001&-5& 56&0.10&1.140 0.011 2&29.39 0.09 1& 31.18 0.12&   15 0.4 1.8& 6.8&0.66&18.3\nl
N4459   &187.250& 13.979& 1553&-1& 56&0.20&1.187 0.015 1&29.47 0.21 1& 31.04 0.22&   37 0.2 1.9& 3.5&1.11&20.6\nl
N4468   &187.380& 14.050& 1232&-3& 56&0.20&1.045 0.015 1&28.79 0.11 1& 31.00 0.14&   19 0.5 2.0& 3.1&1.35&17.4\nl
N4472   &187.444&  7.999& 1346&-5& 56&0.10&1.218 0.011 3&29.62 0.07 2& 31.06 0.10&   65 0.2 2.7& 3.8&1.08&22.5\nl
N4473   &187.453& 13.430& 2575&-5& 56&0.12&1.158 0.012 2&29.28 0.11 1& 30.98 0.13&   35 0.3 2.2& 2.5&1.62&20.3\nl
N4476   &187.495& 12.348& 2296&-3& 56&0.12&1.048 0.017 1&28.98 0.14 1& 31.18 0.17&   20 0.6 2.2& 2.1&1.40&17.8\nl
N4478   &187.572& 12.329& 1711&-5& 56&0.11&1.164 0.019 3&29.62 0.26 1& 31.29 0.28&   19 0.6 2.4& 3.0&1.49&19.3\nl
N4486   &187.707& 12.390& 1632&-4& 56&0.10&1.244 0.012 3&29.72 0.14 2& 31.03 0.16&   53 0.1 3.3& 4.4&0.74&22.2\nl
N4489   &187.718& 16.759& 1290&-5& 56&0.12&1.046 0.015 1&29.05 0.13 1& 31.26 0.15&   15 0.6 1.3& 3.3&1.48&17.9\nl
N4494   &187.851& 25.774& 1653&-5&235&0.09&1.139 0.010 3&29.38 0.08 1& 31.16 0.11&   27 0.2 2.5& 3.2&0.91&20.6\nl
N4526   &188.512&  7.700&  949&-2& 56&0.10&1.188 0.021 4&29.57 0.17 2& 31.14 0.20&   46 0.2 3.8& 5.3&0.92&21.1\nl
N4531   &188.567& 13.076&  345&-1& 56&0.18&1.100 0.015 1&28.94 0.20 1& 30.91 0.22&   42 0.4 1.7& 3.2&1.19&19.4\nl
N4548   &188.860& 14.497&  884& 3& 56&0.16&1.148 0.019 1&29.67 0.53 1& 31.42 0.54&   24 0.4 1.5& 3.1&1.19&22.0\nl
N4546   &188.873& -3.794& 1395&-3&  0&0.15&1.155 0.013 2&29.03 0.19 1& 30.74 0.20&   34 0.3 2.1& 2.9&1.20&19.7\nl
N4550   &188.879& 12.221&  719&-2& 56&0.17&1.078 0.011 2&28.94 0.18 1& 31.00 0.20&   22 0.4 1.6& 3.5&1.11&18.4\nl
U07767  &188.886& 73.675& 1390&-5&  0&0.09&1.152 0.019 1&30.48 0.96 1& 32.22 0.97&   18 0.6 1.3& 2.2&1.67&18.3\nl
N4551   &188.909& 12.266& 1536&-5& 56&0.17&1.170 0.009 3&29.54 0.16 2& 31.19 0.17&   22 0.4 1.7& 3.8&1.05&18.9\nl
N4552   &188.916& 12.557&  660&-5& 56&0.18&1.194 0.015 1&29.39 0.11 1& 30.93 0.14&   59 0.3 2.6& 3.1&1.35&20.5\nl
N4565   &189.086& 25.989& 1527& 3&235&0.07&1.128 0.027 1&29.37 0.11 1& 31.21 0.17&   54 0.6 1.2& 3.2&0.99&21.1\nl
N4564   &189.113& 11.439& 1504&-5& 56&0.15&1.161 0.009 3&29.19 0.16 2& 30.88 0.17&   26 0.3 2.7& 3.2&1.12&19.3\nl
N4589   &189.357& 74.195& 2040&-5& 98&0.12&1.180 0.015 1&30.10 0.20 1& 31.71 0.22&   28 0.3 2.5& 1.2&1.98&20.5\nl
N4578   &189.378&  9.555& 2626&-2& 56&0.09&1.127 0.015 1&29.50 0.10 1& 31.34 0.13&   26 0.4 3.0& 3.6&1.12&19.7\nl
E322-038&189.576&-41.501& 3428&-3& 58&0.56&1.242 0.051 1&31.15 0.25 1& 32.48 0.34&   11 0.7 1.4& 0.2&2.37&20.1\nl
N4594   &189.997&-11.623& 1483& 1&  0&0.22&1.175 0.031 2&28.32 0.10 1& 29.95 0.18&   48 0.7 1.4& 2.6&1.04&21.6\nl
N4600   &190.094&  3.120& 1137&-2&  0&0.12&1.141 0.017 1&27.55 0.20 1& 29.33 0.22&   30 0.7 1.5& 1.7&1.66&16.0\nl
I3653   &190.316& 11.387&  940& 0& 56&0.14&1.139 0.017 1&29.05 0.44 1& 30.84 0.45&   11 0.5 2.0& 2.8&1.28&16.5\nl
N4620   &190.499& 12.943& 1511&-2& 56&0.13&1.048 0.019 1&29.44 0.29 1& 31.64 0.30&   16 0.7 1.4& 2.3&1.51&18.0\nl
N4621   &190.510& 11.647&  780&-5& 56&0.14&1.172 0.018 2&29.67 0.18 1& 31.31 0.20&   48 0.2 3.2& 2.5&1.62&21.4\nl
N4616   &190.571&-40.642& 4880&-4& 59&0.55&1.196 0.021 2&31.45 0.21 1& 32.98 0.24&   17 0.4 1.8& 1.4&2.36&20.3\nl
N4638   &190.699& 11.443& 1484&-3& 56&0.11&1.149 0.013 2&29.93 0.25 1& 31.68 0.26&   26 0.7 1.4& 3.0&1.23&19.8\nl
N4636   &190.707&  2.688& 1286&-5&152&0.12&1.233 0.012 2&29.46 0.11 1& 30.83 0.13&   43 0.2 3.6& 3.3&1.29&21.7\nl
N4649   &190.918& 11.549& 1430&-5& 56&0.12&1.232 0.023 2&29.76 0.09 1& 31.13 0.15&   35 0.2 1.9& 3.8&0.86&22.4\nl
N4645   &191.041&-41.750& 2883&-4& 58&0.64&1.188 0.025 2&30.82 0.13 1& 32.38 0.18&   17 0.3 2.5& 1.9&2.13&20.6\nl
N4660   &191.135& 11.191& 1451&-5& 56&0.14&1.154 0.015 1&28.82 0.17 1& 30.54 0.19&   26 0.4 1.7& 2.3&1.67&18.6\nl
N4684   &191.824& -2.727& 1940&-1&151&0.12&1.100 0.015 1&28.68 0.17 1& 30.65 0.19&   30 0.4 1.5& 0.4&2.68&18.4\nl
E322-088&192.091&-41.714& 2895&-2& 58&0.48&1.156 0.034 1&31.07 0.65 1& 32.78 0.67&   34 0.4 1.8& 1.2&2.46&19.8\nl
N4697   &192.150& -5.801& 1561&-5&151&0.13&1.157 0.010 3&28.64 0.12 1& 30.35 0.14&   41 0.3 2.1& 1.7&2.19&20.8\nl
N4696   &192.208&-41.311& 3248&-4& 58&0.49&1.203 0.014 3&31.25 0.15 2& 32.75 0.17&   57 0.3 2.1& 0.7&2.53&22.4\nl
E322-101&192.391&-41.056& 2343& 0& 58&0.49&1.167 0.023 2&30.52 0.42 1& 32.18 0.44&   10 0.5 2.1& 2.7&1.99&18.3\nl
N4709   &192.517&-41.382& 4939&-5& 59&0.51&1.199 0.011 4&31.22 0.22 1& 32.74 0.23&   26 0.2 3.0& 2.4&1.80&21.9\nl
N4725   &192.612& 25.500& 1502& 2&  0&0.05&1.209 0.023 1&29.14 0.32 1& 30.61 0.34&   27 0.7 1.3& 2.1&1.82&21.8\nl
N4736   &192.723& 41.119&  540& 2&  0&0.08&1.071 0.017 1&26.49 0.15 2& 28.58 0.18&  123 0.6 1.4& 6.2&0.52&19.3\nl
N4733   &192.779& 10.912& 1237&-4& 56&0.09&1.099 0.015 1&28.90 0.18 1& 30.87 0.20&   26 0.3 2.7& 3.9&1.03&18.6\nl
N4742   &192.950&-10.455& 1669&-5&  0&0.18&1.045 0.017 2&28.74 0.13 1& 30.95 0.16&   33 0.5 1.8& 2.1&1.84&18.5\nl
N4754   &193.074& 11.314& 1705&-3& 56&0.14&1.178 0.011 2&29.51 0.12 1& 31.13 0.14&   33 0.3 2.6& 1.7&1.82&20.2\nl
N4753   &193.095& -1.199& 1635& 0&  0&0.15&1.087 0.013 2&29.83 0.17 1& 31.86 0.19&   67 0.7 1.3& 1.1&1.67&21.4\nl
E323-034&193.358&-41.204& 4542&-5& 59&0.54&1.165 0.019 2&31.02 0.20 1& 32.69 0.22&   27 0.5 2.2& 1.1&2.51&20.8\nl
N4802   &193.957&-12.055& 1358&-2&  0&0.21&1.004 0.018 1&27.92 0.13 1& 30.31 0.16&   28 0.7 1.4& 2.2&1.63&17.7\nl
N4826   &194.185& 21.685&  717& 2&  0&0.18&1.029 0.011 2&27.09 0.18 1& 29.37 0.20&  102 0.7 1.4& 5.4&0.77&19.9\nl
N4946   &196.372&-43.591& 3310&-4& 62&0.47&1.190 0.021 2&31.83 0.14 2& 33.39 0.17&   15 0.3 2.8& 1.9&2.45&21.3\nl
N5044   &198.850&-16.386& 3033&-5& 63&0.30&1.210 0.027 1&31.00 0.25 1& 32.47 0.28&   28 0.3 2.4& 0.3&2.18&21.2\nl
N5102   &200.491&-36.630&  706&-3&226&0.24&0.976 0.016 1&25.49 0.09 1& 28.01 0.13&   57 0.3 2.1& 5.3&0.90&17.6\nl
N5128   &201.371&-43.017&  812&-2&226&0.50&1.078 0.016 1&26.05 0.11 1& 28.12 0.14&  185 0.6 1.3& 4.4&1.07&21.8\nl
N5195   &202.495& 47.272&  742& 0&  0&0.16&1.056 0.019 1&27.25 0.25 2& 29.42 0.27&   38 0.0 1.9& 6.0&0.63&19.4\nl
N5273   &205.535& 35.653& 1313&-2&  0&0.04&1.142 0.017 1&29.31 0.24 1& 31.09 0.26&   26 0.4 1.7& 2.2&1.38&18.9\nl
N5322   &207.315& 60.191& 1916&-5&254&0.06&1.183 0.011 2&30.88 0.22 1& 32.47 0.23&   29 0.4 3.0& 1.0&1.74&22.0\nl
N5338   &208.360&  5.208& 1068&-2&  0&0.12&1.019 0.023 1&28.21 0.28 1& 30.54 0.30&   14 0.5 1.1& 3.0&2.05&16.8\nl
N5485   &211.798& 55.002& 2111&-2&237&0.07&1.180 0.017 1&30.46 0.33 1& 32.07 0.34&   32 0.7 1.4& 0.6&2.32&20.5\nl
N5582   &215.181& 39.693& 1492&-5&  0&0.06&1.145 0.013 2&30.51 0.22 1& 32.27 0.23&   24 0.5 1.8& 3.8&1.21&20.1\nl
N5574   &215.235&  3.239& 1845&-3&  0&0.14&1.054 0.011 3&29.71 0.61 1& 31.89 0.62&   17 0.6 1.3& 1.2&2.18&18.4\nl
N5576   &215.268&  3.271& 1818&-5& 68&0.14&1.098 0.010 4&30.05 0.12 1& 32.03 0.14&   24 0.3 2.4& 2.7&1.71&20.8\nl
N5611   &216.022& 33.049& 2154&-2&  0&0.05&1.111 0.017 2&30.09 0.45 1& 32.01 0.46&   15 0.7 1.4& 0.8&2.37&18.4\nl
N5631   &216.639& 56.583& 2057&-2&  0&0.09&1.122 0.019 1&30.35 0.22 1& 32.22 0.24&   32 0.7 1.4& 1.1&2.41&20.3\nl
N5638   &217.419&  3.234& 1901&-5& 68&0.14&1.169 0.011 3&30.44 0.23 2& 32.10 0.24&   25 0.2 3.5& 2.0&1.79&20.5\nl
N5687   &218.722& 54.476& 2310&-3&  0&0.05&1.174 0.013 2&30.60 0.50 1& 32.23 0.51&   16 0.7 1.4& 0.2&2.68&20.2\nl
N5770   &223.313&  3.960& 1687&-2&  0&0.17&1.120 0.017 1&29.52 0.28 1& 31.39 0.29&   32 0.7 1.4& 1.7&1.97&18.8\nl
N5812   &225.232& -7.458& 2286&-5&  0&0.38&1.213 0.015 1&30.70 0.28 1& 32.15 0.29&   33 0.3 2.6& 0.8&2.35&21.2\nl
N5813   &225.297&  1.702& 2178&-5& 70&0.25&1.189 0.014 2&30.97 0.16 1& 32.54 0.18&   25 0.3 2.7& 2.4&1.20&21.9\nl
N5831   &226.030&  1.221& 1894&-5& 70&0.26&1.140 0.010 5&30.38 0.15 1& 32.17 0.17&   18 0.3 2.1& 3.3&1.14&20.3\nl
N5839   &226.367&  1.635& 1420&-2& 70&0.23&1.190 0.011 2&30.21 0.30 1& 31.77 0.30&   32 0.7 1.4& 4.4&1.24&19.2\nl
N5845   &226.505&  1.635& 1654&-5& 70&0.23&1.124 0.012 3&30.21 0.19 1& 32.07 0.21&   14 0.6 2.4& 3.5&0.83&18.8\nl
N5846   &226.622&  1.607& 1917&-5& 70&0.24&1.227 0.007 8&30.59 0.19 1& 31.98 0.20&   23 0.2 3.3& 4.2&1.02&22.0\nl
N5866   &226.626& 55.763&  754&-1&  0&0.06&1.121 0.009 4&29.05 0.10 3& 30.93 0.12&   38 0.6 2.3& 5.8&0.87&20.4\nl
N5898   &229.555&-24.097& 2284&-5& 71&0.63&1.169 0.009 5&30.67 0.25 1& 32.32 0.26&   33 0.4 1.8& 2.9&2.12&21.1\nl
N5903   &229.651&-24.068& 2711&-5& 71&0.64&1.142 0.011 4&30.87 0.22 1& 32.65 0.23&   24 0.3 2.4& 1.3&2.28&21.5\nl
I1153   &239.265& 48.168&  803&-2&  0&0.08&1.197 0.011 3&29.75 0.19 2& 31.28 0.21&   15 0.4 2.9& 5.2&0.83&18.4\nl
N6017   &239.314&  5.998& 1921& 0&  0&0.23&1.092 0.011 2&30.37 0.31 1& 32.37 0.31&   16 0.7 1.4& 1.3&1.96&18.5\nl
N6548   &271.496& 18.587& 2088&-2&  0&0.35&1.244 0.009 3&30.50 0.20 2& 31.81 0.21&   34 0.6 2.6& 1.9&1.77&21.1\nl
N6673   &281.279&-62.297& 1128&-4& 78&0.46&1.189 0.019 1&28.61 0.20 1& 30.18 0.23&   29 0.6 1.3& 3.2&1.38&18.1\nl
N6703   &281.829& 45.551& 2244&-3&  0&0.38&1.164 0.006 9&30.45 0.29 3& 32.13 0.29&   20 0.5 2.0& 1.3&2.02&21.0\nl
N6684   &282.240&-65.174&  846&-2&  0&0.29&1.116 0.015 2&28.82 0.22 1& 30.72 0.24&   72 0.5 1.1& 4.8&1.07&20.1\nl
I4797   &284.122&-54.306& 2544&-4&279&0.34&1.184 0.018 1&30.66 0.18 1& 32.24 0.20&   23 0.4 1.8& 2.1&1.96&20.6\nl
I4889   &296.316&-54.344& 2393&-5& 77&0.23&1.137 0.016 2&30.54 0.13 1& 32.33 0.16&   20 0.4 3.0& 2.2&1.77&20.7\nl
N6869   &300.150& 66.217& 2603&-2&  0&0.79&1.164 0.007 5&30.86 0.21 1& 32.53 0.22&   20 0.5 2.0& 1.4&2.34&20.7\nl
N6851   &300.890&-48.284& 2914&-5& 80&0.20&1.137 0.016 3&30.99 0.19 1& 32.79 0.21&   20 0.4 3.0& 0.7&2.04&20.4\nl
N6861   &301.830&-48.370& 2687&-3& 80&0.23&1.221 0.019 2&30.82 0.35 1& 32.24 0.36&   32 0.5 1.9& 1.2&2.47&21.2\nl
N6909   &306.911&-47.026& 2600&-4& 80&0.16&1.064 0.020 2&30.63 0.12 2& 32.75 0.16&   19 0.4 3.0& 1.9&2.00&19.9\nl
N7029   &317.968&-49.283& 2645&-5& 83&0.16&1.138 0.032 1&31.13 0.46 1& 32.92 0.49&   32 0.5 1.9& 1.4&2.27&20.6\nl
N7041   &319.136&-48.364& 1697&-3& 83&0.17&1.139 0.055 2&30.27 0.18 1& 32.05 0.31&   37 0.4 1.6& 3.2&1.66&20.2\nl
N7049   &319.751&-48.564& 1977&-2& 83&0.24&1.174 0.033 1&30.75 0.15 1& 32.38 0.22&   55 0.5 2.1& 2.8&1.58&21.8\nl
N7097   &325.056&-42.540& 2184&-5&267&0.09&1.176 0.022 2&30.93 0.18 1& 32.55 0.21&   29 0.2 2.0& 3.7&1.55&20.2\nl
N7144   &328.179&-48.254& 1722&-5& 84&0.09&1.161 0.009 5&30.26 0.10 3& 31.95 0.12&   33 0.2 1.9& 3.1&1.65&20.7\nl
N7145   &328.334&-47.882& 1673&-5& 84&0.09&1.133 0.016 2&30.04 0.19 1& 31.85 0.21&   25 0.3 2.4& 2.5&1.76&20.1\nl
N7173   &330.515&-31.974& 2225&-4& 85&0.11&1.150 0.017 1&30.74 0.16 1& 32.48 0.19&   21 0.3 2.8& 2.8&1.89&19.9\nl
N7168   &330.530&-51.742& 2563&-5&102&0.10&1.184 0.022 2&31.14 0.21 1& 32.72 0.24&   29 0.2 2.0& 1.8&2.36&20.4\nl
N7180   &330.576&-20.547& 1164&-2&265&0.14&1.109 0.009 4&29.34 0.25 2& 31.26 0.26&   23 0.4 1.7& 3.7&1.35&17.8\nl
N7185   &330.734&-20.471& 1523&-3&265&0.14&1.080 0.012 2&29.45 0.36 1& 31.51 0.37&   30 0.6 1.3& 1.5&2.18&18.5\nl
N7196   &331.478&-50.120& 2813&-5&102&0.09&1.211 0.022 2&31.80 0.28 1& 33.27 0.30&   31 0.2 1.9& 2.5&1.86&21.7\nl
N7192   &331.708&-64.316& 2764&-4&  0&0.15&1.174 0.032 1&31.26 0.28 1& 32.89 0.32&   27 0.3 2.2& 1.0&2.60&21.2\nl
N7200   &331.788&-49.996& 2702&-4&102&0.08&1.183 0.018 3&31.12 0.31 1& 32.71 0.32&   16 0.5 1.9& 2.6&2.03&19.2\nl
N7280   &336.615& 16.149& 1542&-1&  0&0.24&1.105 0.009 3&29.98 0.21 1& 31.93 0.22&   32 0.7 1.4& 2.7&1.67&19.1\nl
N7302   &338.100&-14.117& 2242&-3&  0&0.30&1.126 0.016 3&30.03 0.19 1& 31.88 0.21&   17 0.6 1.3& 1.0&2.33&19.2\nl
N7331   &339.272& 34.419&  492& 3&  0&0.39&1.120 0.017 1&28.72 0.14 1& 30.59 0.17&  115 0.8 1.5& 9.3&0.36&20.9\nl
N7332   &339.352& 23.798&  853&-2&  0&0.16&1.107 0.008 4&29.88 0.19 3& 31.81 0.20&   27 0.6 1.6& 7.5&0.61&20.1\nl
I1459   &344.290&-36.460& 1350&-5&231&0.07&1.194 0.018 1&30.79 0.26 1& 32.33 0.28&   54 0.3 2.2& 4.1&1.49&22.2\nl
N7457   &345.250& 30.144&  478&-3&  0&0.23&1.104 0.009 3&28.66 0.20 3& 30.61 0.21&   30 0.2 2.6& 9.3&0.36&18.8\nl
N7454   &345.278& 16.390& 1637&-5&232&0.34&1.123 0.012 4&30.03 0.29 2& 31.89 0.30&   23 0.5 1.9& 3.5&1.23&19.6\nl
N7507   &348.032&-28.541& 1240&-5&243&0.21&1.196 0.009 4&30.45 0.15 1& 31.99 0.17&   35 0.2 1.7& 3.9&1.20&21.8\nl
N7619   &350.061&  8.206& 3370&-5& 87&0.35&1.229 0.009 3&32.23 0.31 2& 33.62 0.31&   25 0.4 3.5& 2.1&2.46&22.9\nl
I5328   &353.321&-45.017& 2902&-5&  0&0.06&1.173 0.023 2&31.26 0.10 2& 32.89 0.15&   23 0.3 2.5& 1.6&2.67&21.2\nl
N7743   &356.090&  9.934& 1283&-1&  0&0.31&1.080 0.009 3&29.53 0.15 1& 31.58 0.17&   14 0.4 1.6& 4.1&1.03&19.6\nl
N7796   &359.749&-55.457& 3078&-4&  0&0.04&1.230 0.022 2&32.11 0.38 1& 33.49 0.40&   23 0.3 2.5& 0.8&2.43&22.1\nl
\enddata
\label{tab:1}
\end{deluxetable}

\clearpage
\setlength{\tabcolsep}{4pt}
\begin{deluxetable}{lrrrrrrrrrrrrrrr}
\tablewidth{0pt}
\tablecaption{Uncertain SBF Data}
\scriptsize
\tablehead{
\colhead{Galaxy}   & \colhead{RA} & \colhead{Dec}    & \colhead{$v_{CMB}$} &
 \colhead{T} & \colhead{Grp}   & \colhead{$A_B$} & 
\colhead{$(V{-}I)$} & \colhead{$\overline m_I$}   & \colhead{$(m{-}M)$}   &
\colhead{$\langle r \rangle$} &
\colhead{$Q$}   & \colhead{$PD$} & \colhead{$\overline N_I$}
}
\startdata
N1331   & 51.618&-21.356& 1173&-5& 31&0.13&            0&29.73 0.32 1& 31.80 0.37&   21 0.7 1.4& 3.6&1.65&17.6\nl
N1521   & 62.079&-21.052& 4070&-5&  0&0.18&1.152 0.029 1&32.41 0.36 1& 34.14 0.38&   21 0.3 2.8& 0.2&2.81&22.1\nl
N1700   & 74.234& -4.868& 3844&-5&100&0.19&1.163 0.011 4&31.54 0.14 1& 33.23 0.16&   26 0.3 2.2& 0.6&2.88&21.7\nl
N3413   &162.838& 32.768&  917&-2&  0&0.10&0.733 0.015 1&27.44 0.23 1& 31.05 0.24&   18 0.5 1.9& 4.7&0.78&16.2\nl
N3522   &166.669& 20.086& 1590&-5&  0&0.10&            0&30.02 0.21 1& 32.03 0.28&   18 0.5 2.0& 3.2&1.30&18.1\nl
N3962   &178.666&-13.973& 2192&-5&244&0.20&1.145 0.023 1&30.98 0.48 1& 32.74 0.49&   36 0.4 1.7& 0.8&2.85&21.8\nl
N4168   &183.069& 13.207& 2653&-5&  0&0.16&1.132 0.015 1&30.63 0.41 1& 32.45 0.42&   19 0.6 2.3&-0.4&2.31&20.8\nl
N4233   &184.277&  7.623& 2726&-2&150&0.10&1.191 0.015 1&31.09 0.93 1& 32.65 0.94&   32 0.7 1.4&-0.6&2.43&20.3\nl
N4281   &185.090&  5.387& 3089&-1&150&0.09&1.167 0.015 1&30.28 0.26 1& 31.94 0.28&   27 0.3 2.6&-0.1&2.22&20.3\nl
N4627   &190.499& 32.575& 1102&-5&248&0.07&0.911 0.015 1&27.04 0.13 1& 29.86 0.15&   27 0.2 1.6& 3.4&1.32&15.4\nl
N4767   &193.472&-39.714& 3298&-5& 58&0.46&1.175 0.028 2&30.95 0.21 1& 32.58 0.25&   19 0.3 2.2& 1.0&2.80&20.8\nl
N5011   &198.216&-43.097& 3371&-5& 62&0.43&1.189 0.027 1&31.54 0.18 1& 33.11 0.22&   22 0.3 2.7& 0.9&2.73&22.0\nl
N5193   &202.973&-33.235& 3932&-5& 64&0.24&1.164 0.034 1&30.99 0.25 2& 32.66 0.29&   20 0.3 3.0& 0.0&2.91&20.6\nl
N5869   &227.455&  0.470& 2314&-2& 70&0.23&1.161 0.014 5&30.28 1.22 1& 31.98 1.22&    8 0.7 1.3&-0.1&2.24&19.7\nl
N6702   &281.741& 45.707& 4600&-5&  0&0.47&1.149 0.006 8&31.76 0.23 1& 33.51 0.24&   13 0.4 2.9& 0.4&2.71&21.1\nl
I4943   &301.618&-48.374& 2799&-5& 80&0.22&1.133 0.023 2&30.67 0.15 2& 32.48 0.19&   15 0.5 1.9& 1.4&2.80&19.2\nl
N6868   &302.474&-48.379& 2742&-5& 80&0.24&1.230 0.009 4&30.76 0.24 1& 32.14 0.25&   26 0.3 2.3& 0.3&2.77&21.4\nl
I5269   &344.425&-36.017& 1846&-2&231&0.07&            0&30.15 0.24 1& 32.02 0.30&   22 0.7 1.4& 1.8&2.40&19.1\nl
N7562   &348.989&  6.688& 3231&-5& 87&0.45&1.187 0.018 1&32.22 0.83 1& 33.79 0.84&   32 0.7 1.4& 0.2&2.97&22.0\nl
\enddata
\label{tab:2}
\end{deluxetable}

\clearpage
\begin{deluxetable}{lrrrrrrrrrrrrrrr}
\tablewidth{0pt}
\tablecaption{SBF Color Data}
\tablehead{
\colhead{Galaxy}   & 
\colhead{$\langle r\rangle$}   & 
\colhead{$\overline m_V$} &
\colhead{$\overline m_R$} &
\colhead{$\overline m_I$} &
\colhead{$(V{-}I)$} &
\colhead{$\overline m_V{-}\overline m_R$} &
\colhead{$\overline m_V{-}\overline m_I$}
}
\startdata
 N0147&    &  24.51 0.09&  23.39 0.09&  22.29 0.09&  1.030&   1.12 0.12&  2.22 0.12 \nl
 N0185&    &  24.38 0.11&  23.21 0.09&  21.98 0.08&  1.023&   1.18 0.14&  2.40 0.14 \nl
 N0205&  8 &  23.72 0.17&  23.22 0.18&  22.26 0.17&  0.845&   0.50 0.24&  1.46 0.24 \nl
 N0205& 16 &  24.32 0.16&  23.53 0.10&  22.28 0.08&  0.895&   0.79 0.19&  2.04 0.18 \nl
 N0205& 33 &  24.60 0.09&  23.60 0.09&  22.35 0.08&  0.965&   1.00 0.12&  2.25 0.12 \nl
 N0205& 66 &  24.77 0.08&  23.67 0.08&  22.43 0.08&  0.995&   1.10 0.11&  2.34 0.11 \nl
 N0205&132 &  24.77 0.08&  23.67 0.08&  22.45 0.08&  0.985&   1.10 0.11&  2.32 0.11 \nl
 N0221&    &  25.43 0.09&  24.20 0.09&  22.84 0.09&  1.139&   1.22 0.12&  2.59 0.12 \nl
 N0224&    &  25.51 0.09&  24.53 0.09&  22.96 0.09&  1.233&   0.98 0.12&  2.54 0.12 \nl
 N0404&    &  28.13 0.13&  26.99 0.09&  25.48 0.09&  1.063&   1.14 0.16&  2.65 0.16 \nl
\enddata
\label{tab:3}
\end{deluxetable}

\clearpage
\begin{deluxetable}{lrrrrrrrrrrrrrrr}
\tablewidth{0pt}
\tablecaption{SBF Groups}
\tablehead{
\colhead{Group}   & 
\colhead{Example}   & 
\colhead{RA}   & 
\colhead{Dec}   & 
\colhead{$v_h$}   & 
\colhead{7S\#}   & 
\colhead{$(m{-}M)$} &
\colhead{$D$} &
\colhead{$N$}
}
\startdata
LocalGroup & N0224&   10.0&   41.0&$-$300& 282& 24.48 0.05&  0.79    0.02   &  2\nl
Cetus      & N0636&   24.2& $-$7.8&  1800&  26& 31.97 0.09& 24.8\phn 1.1\phn&  5\nl
N1023      & N1023&   37.0&   35.0&   650&   0& 30.00 0.11& 10.0\phn 0.5\phn&  4\nl
Eridanus   & N1407&   53.0&$-$21.0&  1700&  32& 32.00 0.08& 25.1\phn 1.0\phn&  7\nl
Fornax     & N1399&   54.1&$-$35.6&  1400&  31& 31.49 0.04& 19.9\phn 0.4\phn& 26\nl
Dorado     & N1549&   63.7&$-$55.7&  1300& 211& 31.34 0.08& 18.5\phn 0.7\phn&  6\nl
M81        & N3031&  147.9&   69.3& $-$40& 299& 28.01 0.12&  4.0\phn 0.2\phn&  2\nl
LeoIII     & N3193&  153.9&   22.1&  1400&  45& 32.38 0.15& 29.9\phn 2.2\phn&  2\nl
LeoI       & N3379&  161.3&   12.8&   900&  57& 30.22 0.06& 11.1\phn 0.3\phn&  5\nl
LeoII      & N3607&  168.6&   18.3&   950&  48& 31.69 0.08& 21.8\phn 0.9\phn&  5\nl
UMa        & N3928&  180.0&   47.0&   900& 155& 30.84 0.10& 14.7\phn 0.8\phn&  6\nl
N4125      & N4125&  181.4&   65.5&  1300&  54& 31.82 0.17& 23.1\phn 2.0\phn&  3\nl
ComaI      & N4278&  184.4&   29.6&  1000&  55& 31.07 0.10& 16.4\phn 0.8\phn&  4\nl
N4386      & N4386&  185.6&   75.8&  1650&  98& 31.87 0.17& 23.6\phn 2.0\phn&  3\nl
Virgo      & N4486&  187.1&   12.7&  1150&  56& 31.15 0.03& 17.0\phn 0.3\phn& 31\nl
ComaII     & N4494&  187.2&   26.1&  1350& 235& 31.14 0.09& 16.9\phn 0.8\phn&  3\nl
Centaurus  & N4696&  191.5&$-$41.0&  3000&  58& 32.64 0.08& 33.8\phn 1.4\phn&  9\nl
N5011      & N5011&  197.5&$-$42.8&  3100&  62& 33.28 0.14& 45.3\phn 3.1\phn&  2\nl
CenA       & N5128&  200.0&$-$39.0&   550& 226& 28.06 0.11&  4.1\phn 0.2\phn&  2\nl
N5322      & N5322&  212.5&   57.0&  2000& 254& 32.29 0.15& 28.7\phn 2.2\phn&  4\nl
N5846      & N5846&  226.0&    1.8&  1700&  70& 32.17 0.09& 27.1\phn 1.2\phn&  5\nl
Telescopium& N6868&  301.6&$-$48.5&  2900&  80& 32.57 0.10& 32.7\phn 1.6\phn&  5\nl
N7331      & N7457&  338.7&   34.2&   800&   0& 30.60 0.14& 13.2\phn 0.9\phn&  2\nl
\enddata
\label{tab:4}
\end{deluxetable}

\endinsfig  

\clearpage

\manufig{
\centerline{\bf FIGURE CAPTIONS}
\bigskip
}

\manufig{
\figcaption[rc3lookup.eps]{
The cumulative counts of galaxies with $B_T \le 12.5$ are shown for
the RC3 (upper curves) and the SBF survey (lower curves).  The three
sets of curves show counts of E, S0, and Sa and Sb spiral galaxies
with $T\le3$.
\label{fig:complete}}
}

\manufig{
\figcaption[vihisto.eps]{
The \vi\ distribution of SBF survey galaxies, represented by the
shaded histogram.  The dashed curve shows for comparison a Gaussian
of mean $\vi=1.165$ and dispersion 0.04\,mag.
\label{fig:vihisto}}
}

\manufig{
\figcaption[bulkflow.eps]{
The bulk flow observed and predicted in top-hat spheres of radius $R$ is
shown for typical CDM models of the power spectrum (curves), a
simulated universe with a $\Gamma=0.21, \sigma_8=0.94$ CDM power
spectrum (large open points), and for
the SBF-II observations (circles for SBF-II model, diamonds for
Willick and Batra variant).  The simulated universe encompasses many
volumes equal to that of the SBF-II survey, and the rms variation in
bulk flow seen from volume to volume is shown as error bars on the
open points.  Note that the
radius is expressed in terms of $h^{-1}$~Mpc for consistency with what
is normally found in the literature.  The rollover in the data points
at $R<10h^{-1}$~Mpc 
is caused by the limited spatial resolution of the SBF-II model.
\label{fig:bulky}}
}

\manufig{
\figcaption[nbarclr.eps]{
The \vi\ color of SBF galaxies is shown as a function of \Nbar.
\label{fig:nbarclr}}
}

\manufig{
\figcaption[nbar.eps]{
The fluctuation magnitude \mi\ for SBF galaxies in eight selected
groups is shown as a function of $\Nbar_I$.  The lines are {\it not}
fitted to the data, but are drawn with
a slope of 0.14 and zero points according to the SBF distance of the
group derived from \vi\ and \mi.
\label{fig:nbar}}
}

\manufig{
\figcaption[nbarsig.eps]{
The velocity dispersion of elliptical SBF galaxies is shown as a
function of $\Nbar_I$.
\label{fig:nbarsig}}
}

\manufig{
\figcaption[psfactual.eps]{
The number of hours of exposure time is shown as a function of seeing
for the observations which eventually contributed to the tabulated
values for \mi.
\label{fig:psfactual}}
}

\manufig{
\figcaption[psfideal.eps]{
The number of hours of exposure time is shown as a function of seeing
for a new set of observations which would improve the error in \mi\ to 
a median of 0.12 mag.
\label{fig:psfideal}}
}


\begin{figure}[H]
\epsscale{1.0}
\plotone{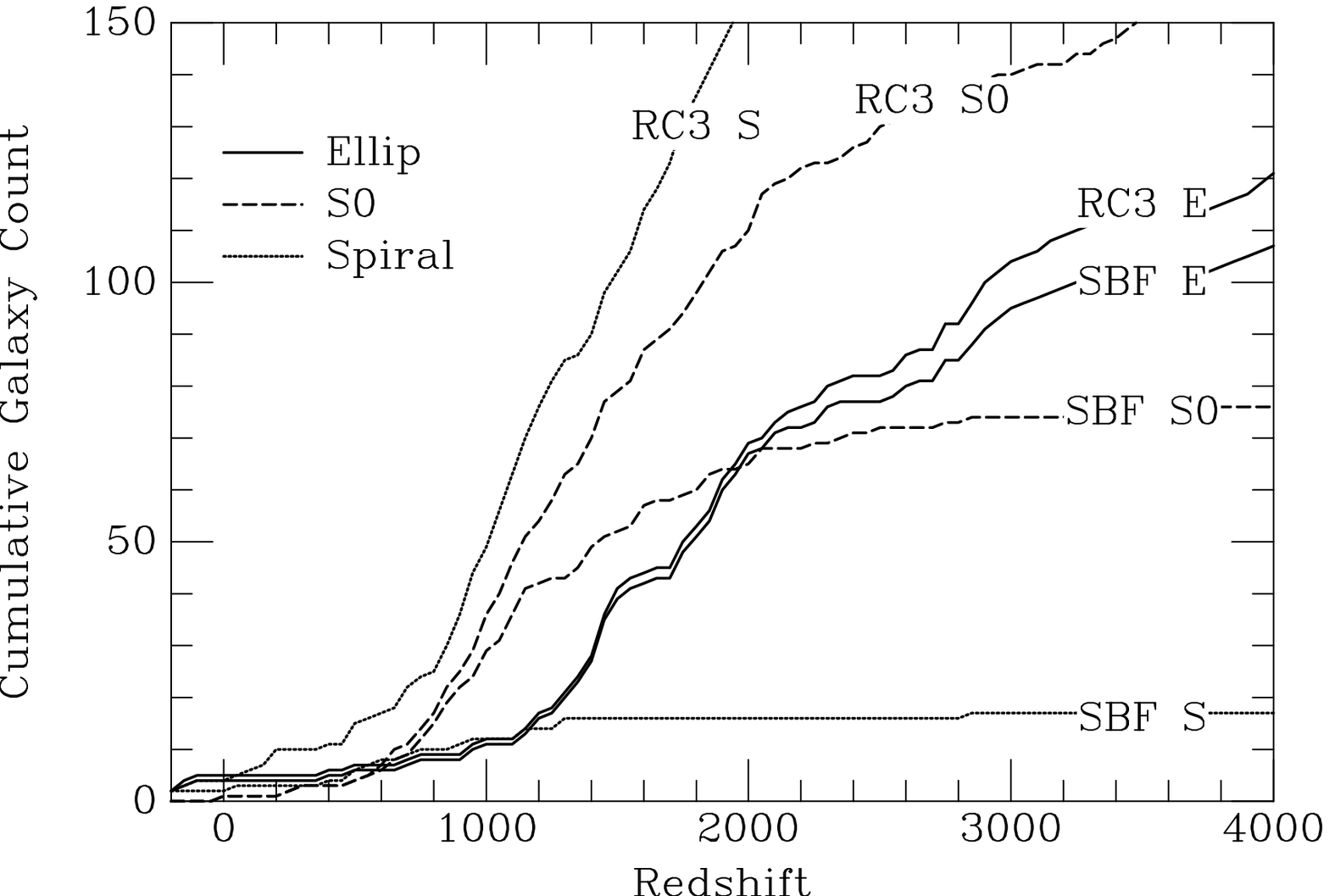}
\capfig{
\caption[rc3lookup.eps]{
The cumulative counts of galaxies with $B_T \le 12.5$ are shown for
the RC3 (upper curves) and the SBF survey (lower curves).  The three
sets of curves show counts of E, S0, and Sa and Sb spiral galaxies
with $T\le3$.
\label{fig:complete}}
}
\end{figure}

\begin{figure}[H]
\epsscale{1.0}
\plotone{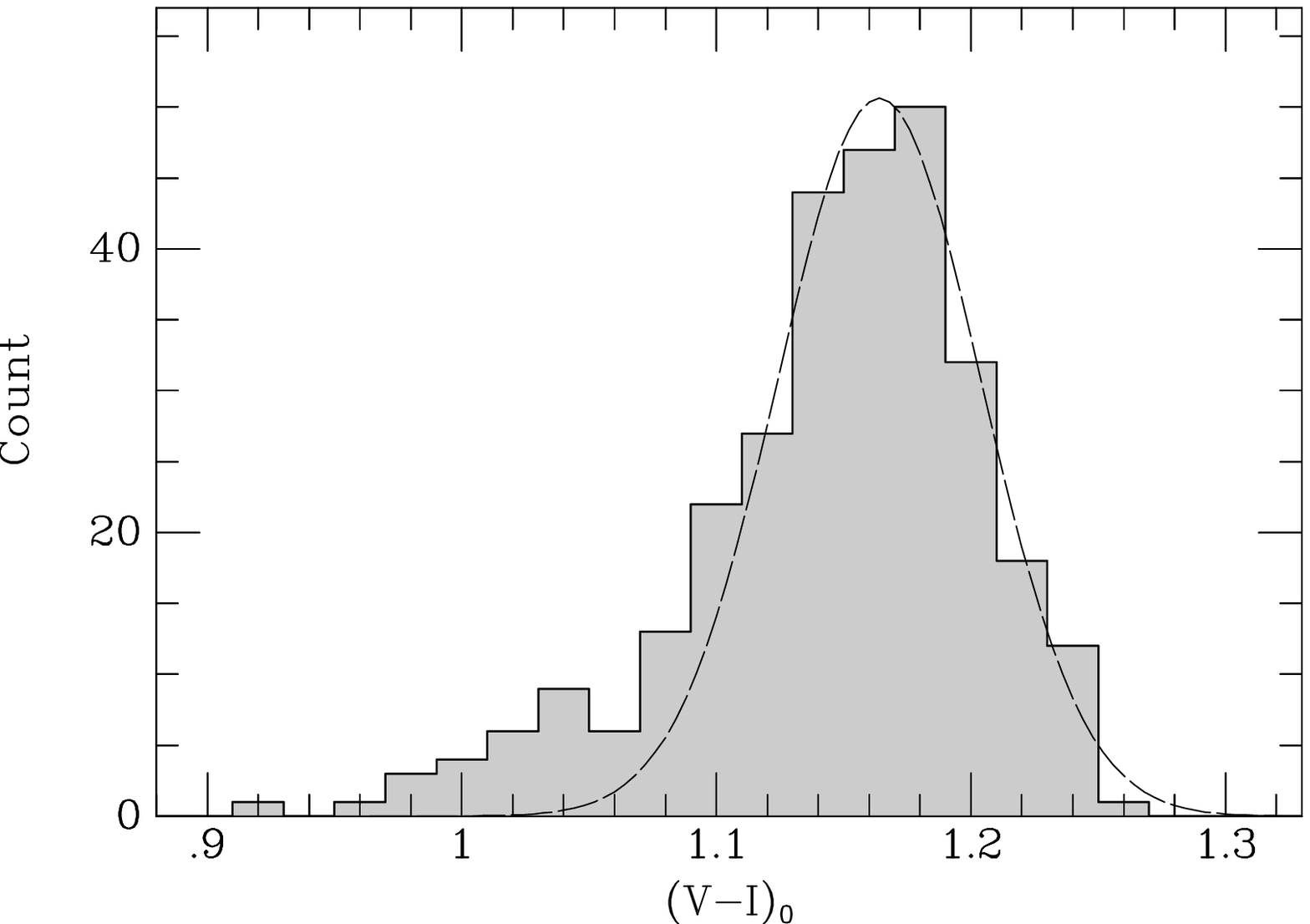}
\capfig{
\caption[vihisto.eps]{
The \vi\ distribution of SBF survey galaxies, represented by the
shaded histogram.  The dashed curve shows for comparison a Gaussian
of mean $\vi=1.165$ and dispersion 0.04\,mag.
\label{fig:vihisto}}
}
\end{figure}

\begin{figure}[H]
\epsscale{1.0}
\plotone{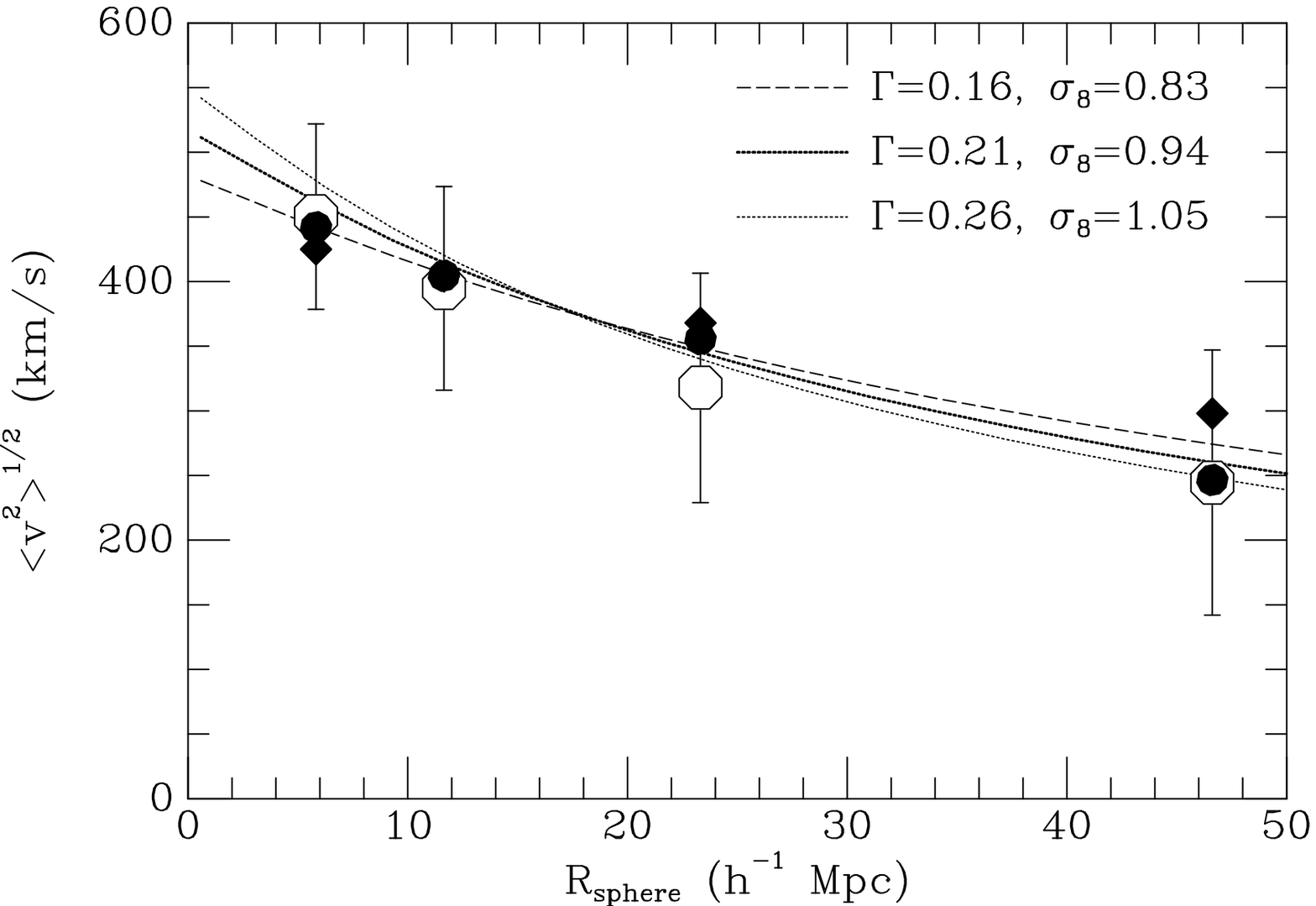}
\capfig{
\caption[bulkflow.eps]{
The bulk flow observed and predicted in top-hat spheres of radius $R$ is
shown for typical CDM models of the power spectrum (curves), a
simulated universe with a $\Gamma=0.21, \sigma_8=0.94$ CDM power
spectrum (large open points), and for
the SBF-II observations (circles for SBF-II model, diamonds for
Willick and Batra model).  The simulated universe encompasses many
volumes equal to that of the SBF-II survey, and the rms variation in
bulk flow seen from volume to volume is shown as error bars on the
open points.  Note that the
radius is expressed in terms of $h^{-1}$~Mpc for consistency with what
is normally found in the literature.  The rollover in the data points
at $R<10h^{-1}$~Mpc 
is caused by the limited spatial resolution of the SBF-II model.
\label{fig:bulky}}
}
\end{figure}

\begin{figure}[H]
\epsscale{1.0}
\plotone{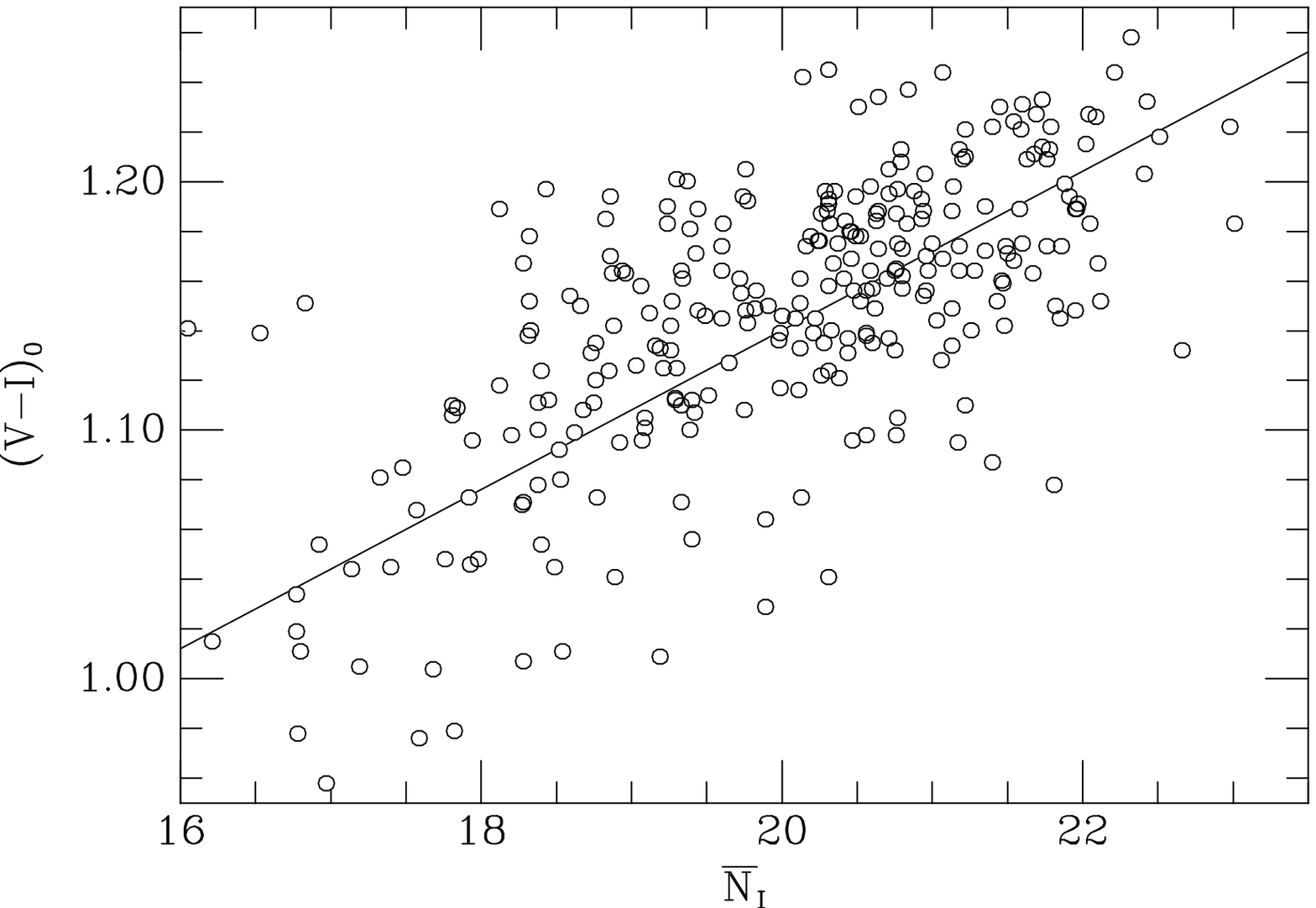}
\capfig{
\caption[nbarclr.eps]{
The \vi\ color of SBF galaxies is shown as a function of \Nbar.
\label{fig:nbarclr}}
}
\end{figure}

\begin{figure}[H]
\epsscale{1.0}
\plotone{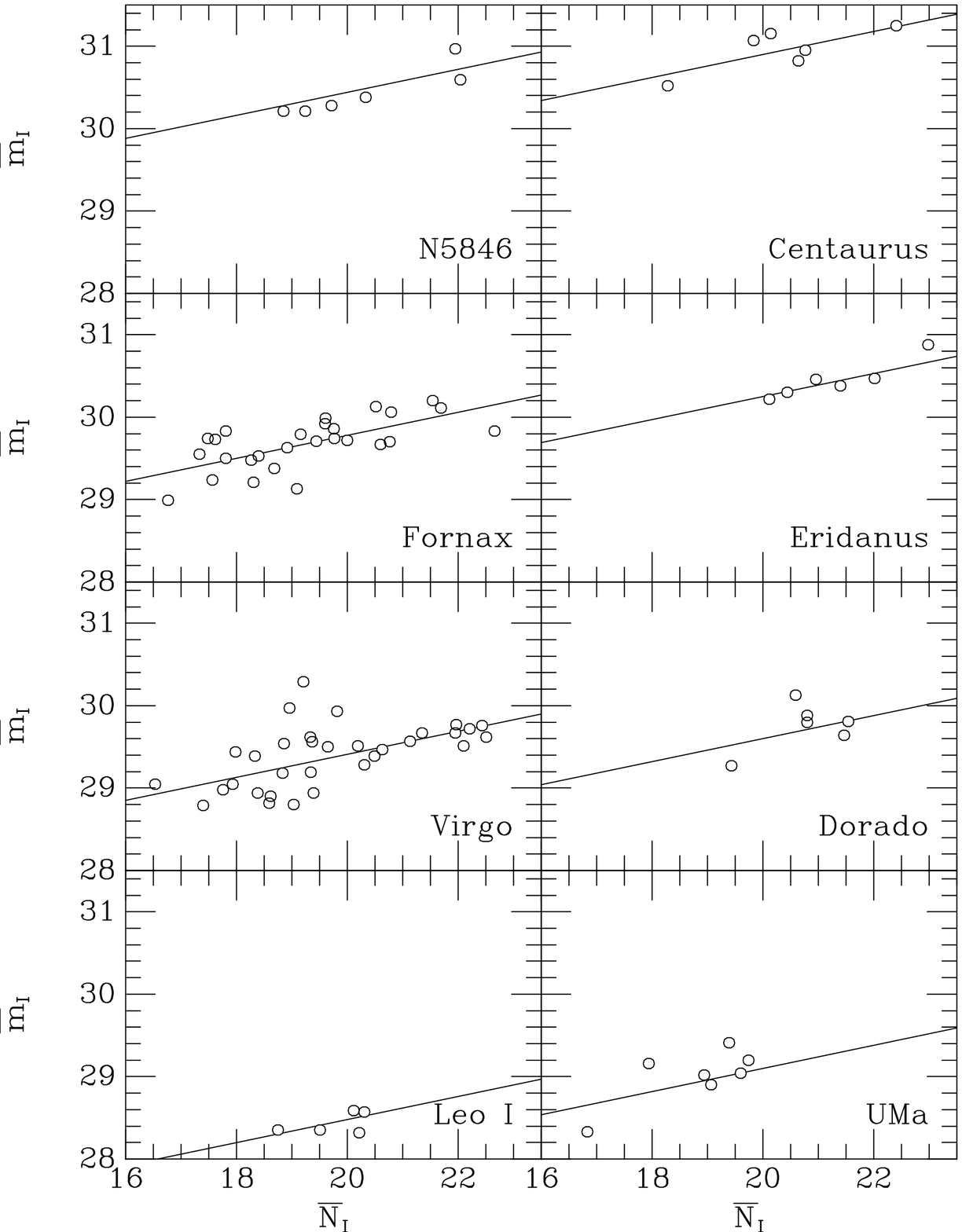}
\capfig{
\caption[nbar.eps]{
The fluctuation magnitude \mi\ for SBF galaxies in eight selected
groups is shown as a function of $\Nbar_I$.  The lines are {\it not}
fitted to the data, but are drawn with
a slope of 0.14 and zero points according to the SBF distance of the
group derived from \vi\ and \mi.
\label{fig:nbar}}
}
\end{figure}

\begin{figure}[H]
\epsscale{1.0}
\plotone{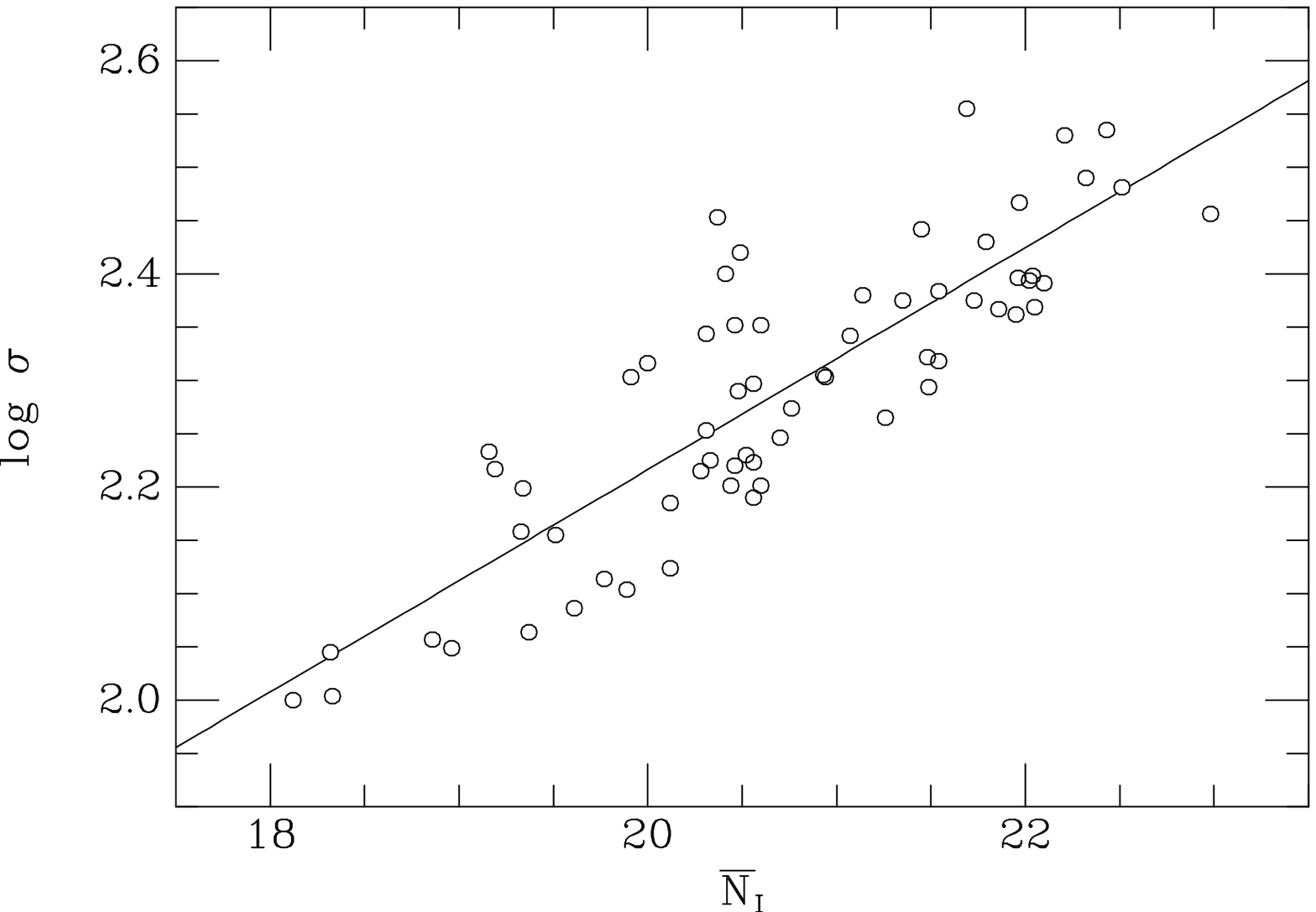}
\capfig{
\caption[nbarsig.eps]{
The velocity dispersion of elliptical SBF galaxies is shown as a
function of $\Nbar_I$.
\label{fig:nbarsig}}
}
\end{figure}

\begin{figure}[H]
\epsscale{1.0}
\plotone{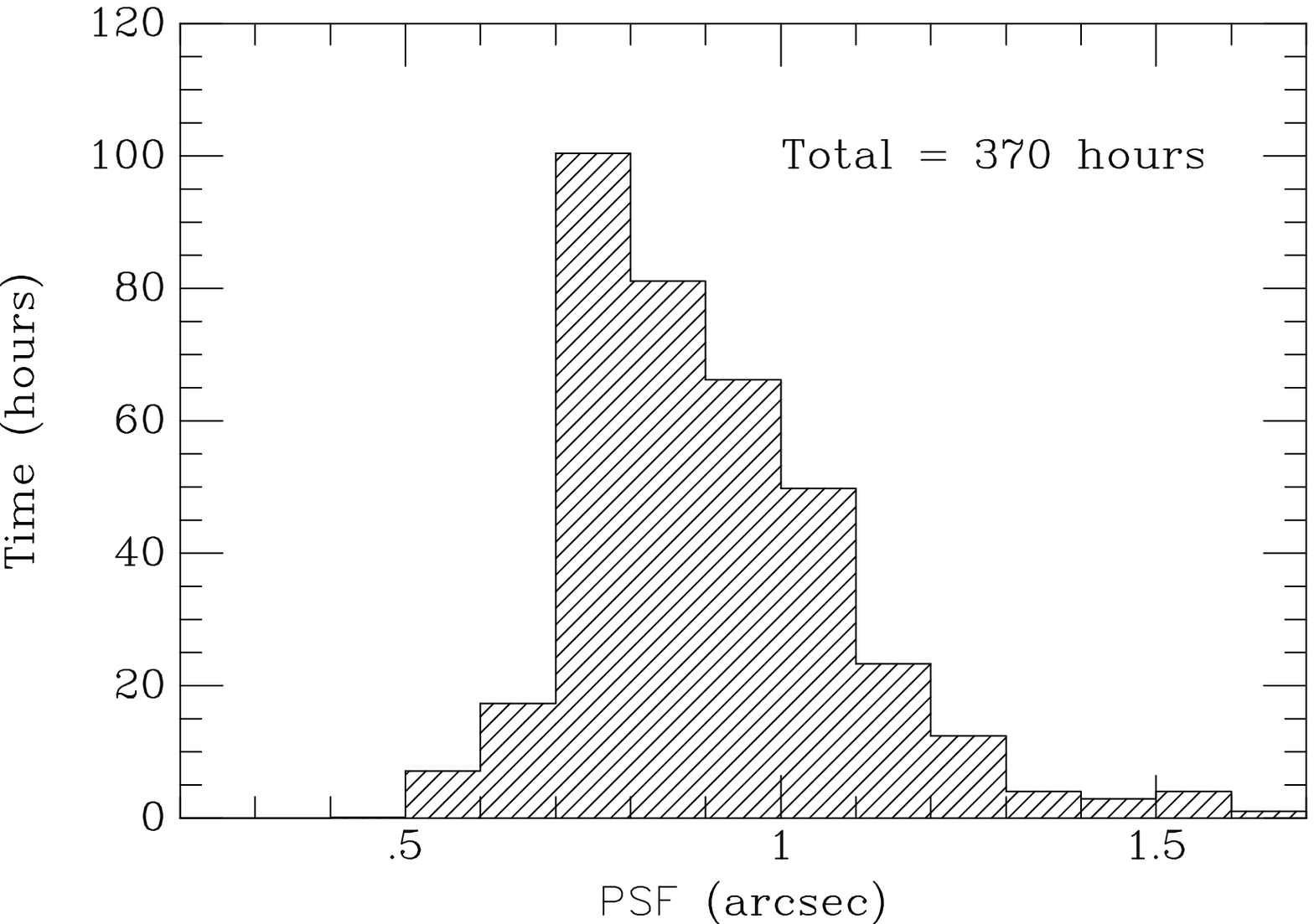}
\capfig{
\caption[psfactual.eps]{
The number of hours of exposure time is shown as a function of seeing
for the observations which eventually contributed to the tabulated
values for \mi.
\label{fig:psfactual}}
}
\end{figure}

\begin{figure}[H]
\epsscale{1.0}
\plotone{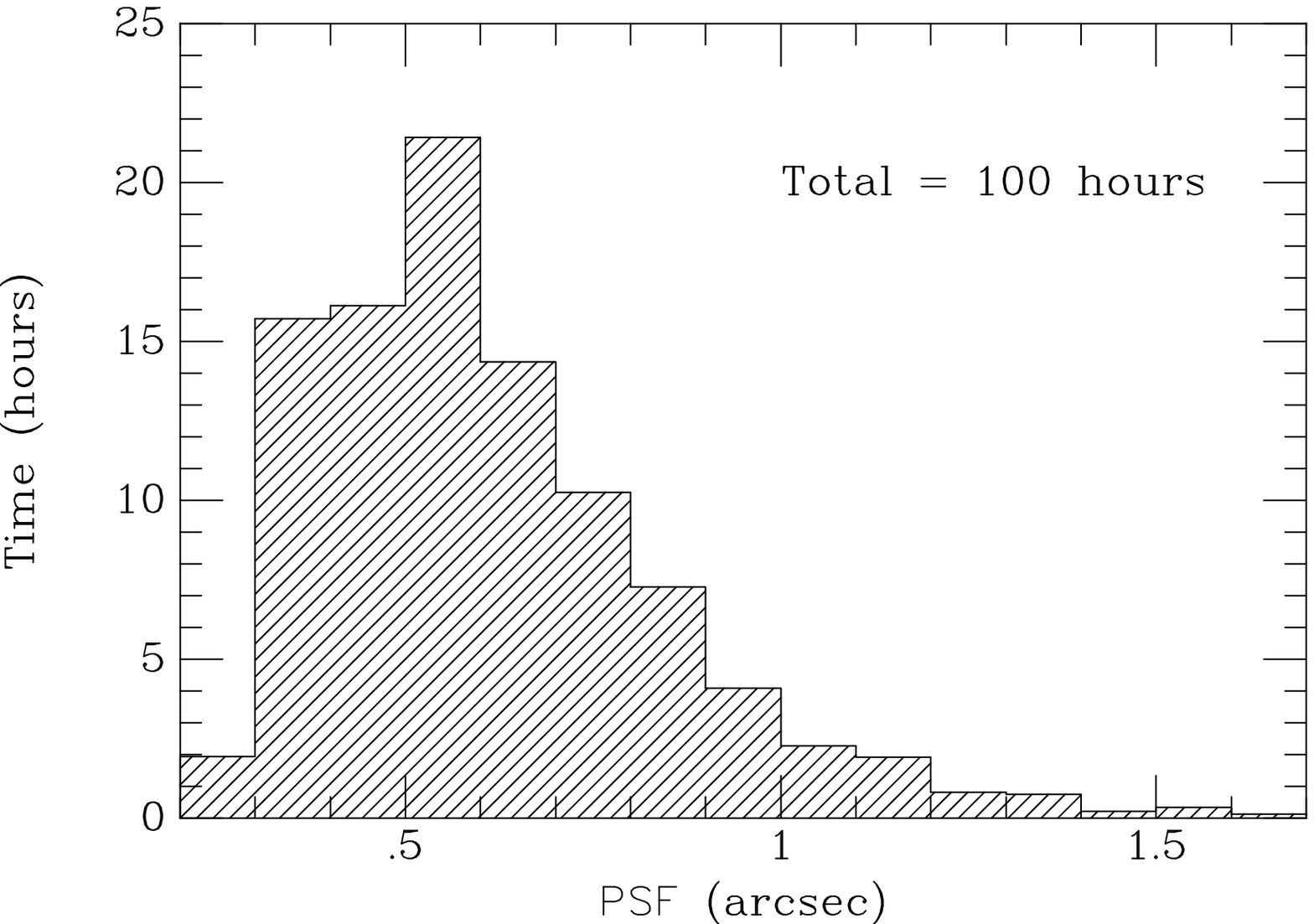}
\capfig{
\caption[psfideal.eps]{
The number of hours of exposure time is shown as a function of seeing
for a new set of observations which would improve the error in \mi\ to 
a median of 0.12 mag.
\label{fig:psfideal}}
}
\end{figure}

\end{document}